\documentclass[usenatbib]{mnras}
\usepackage[T1]{fontenc}
\usepackage{ae,aecompl}
\usepackage{graphics,txfonts}
\usepackage{gensymb}

\usepackage{longtable}
\usepackage{amssymb}
\usepackage{xcolor}
\usepackage{color}
\usepackage{lipsum}
\usepackage{textcomp}
\usepackage{enumerate}
\usepackage{placeins}
\usepackage{float}
\usepackage{blindtext}

\newcommand{\msol}{\ensuremath{M_{\odot}}\xspace}
\newcommand{\rsol}{\ensuremath{R_{\odot}}\xspace}
\newcommand{\lsol}{\ensuremath{L_{\odot}}\xspace}
\newcommand{\rxte}{\textit{RXTE}\xspace}
\newcommand{\nustar}{\textit{NuSTAR}\xspace}
\newcommand{\astrosat}{\textit{Astrosat}\xspace}

\newcommand{\swift}{\textit{Swift}\xspace}
\newcommand{\suzaku}{\textit{Suzaku}\xspace}

\newcommand{\source}{4U~1909+07\xspace}

\newcommand{\kms}{km~s$^{-1}$}

\newcommand{\lumcgs}{erg~s$^{-1}$}
\newcommand{\nh}{$N_{\rm H}$~}

\usepackage{tikz,hyperref}

\title[{NuSTAR}  and {Astrosat} observations of 4U~1909+07]
{Revisiting the spectral and timing properties of 4U~1909+07 with {\em NuSTAR}  and {\em Astrosat}}

\author[Jaisawal et al.]
{Gaurava K. Jaisawal$^{1}$\thanks{E-mail: gaurava@space.dtu.dk}, 
Sachindra Naik$^2$,
Wynn C.G. Ho$^3$,
Neeraj Kumari$^{2,4}$,
\newauthor
Prahlad Epili$^5$,
and Georgios~Vasilopoulos$^6$
\\
$^1$ National Space Institute, Technical University of Denmark, Elektrovej 327-328, DK-2800 Lyngby, Denmark\\ 
$^2$ Astronomy and Astrophysics Division, Physical Research Laboratory, Navrangapura, Ahmedabad - 380009, Gujarat, India\\
$^3$ Department of Physics and Astronomy, Haverford College, 370 Lancaster Avenue, Haverford, PA, 19041, USA \\
$^4$ Indian Institute of Technology, Gandhinagar - 382355, Gujarat, India\\
$^5$ School of Physics and Technology, Wuhan University, Wuhan 430072, China \\
$^6$ Department of Astronomy, Yale University, PO Box 208101, New Haven, CT 06520-8101, USA \\
}

\pubyear{2020}

\begin{document}
\label{firstpage}
\pagerange{\pageref{firstpage}--\pageref{lastpage}}
\maketitle

\begin{abstract}
We present the results obtained from the analysis of high mass X-ray binary 
pulsar 4U~1909+07 using {\em NuSTAR} and {\em Astrosat} observations in 2015 
and 2017 July, respectively. X-ray pulsations at $\approx$604~s are 
clearly detected in our study. Based on the long term spin-frequency evolution, 
the source is found to spun up in the last 17 years. We observed a strongly 
energy-dependent pulse profile that evolved from a complex broad structure in 
soft X-rays into a profile with a narrow emission peak followed by a plateau 
in energy ranges above 20~keV. This behaviour ensured a positive correlation 
between the energy and pulse fraction. The pulse profile morphology and its 
energy-evolution are almost similar during both the observations, suggesting 
a persistent emission geometry of the pulsar over time. The broadband energy 
spectrum of the pulsar is approximated by an absorbed high energy exponential 
cutoff power law model with iron emission lines. In contrast to the previous report, 
we found no statistical evidence for the presence of cyclotron absorption features 
in the X-ray spectra. We performed phase-resolved spectroscopy by using 
data from the {\em NuSTAR} observation. Our results showed a clear signature of 
absorbing material at certain pulse-phases of the pulsar. These findings are 
discussed in terms of stellar wind distribution and its effect on the beam 
geometry of this wind-fed accreting neutron star. We also reviewed the 
subsonic quasi-spherical accretion theory and its implication 
on the magnetic field of \source depending on the global spin-up rate.           

\end{abstract}


\section{INTRODUCTION}
\label{sec:intro}

High mass X-ray binaries (HMXBs) are known to consist of a compact object (mostly a neutron star) 
and a massive ($>$10~\msol) OB optical companion in a close binary  (see, e.g., \citealt{Lewin1997, 
Tauris2006}). The material from the companion primarily gets accreted onto the neutron star by 
falling into its deep gravitation potential. This process releases an enormous amount of energy, mostly 
in the X-ray range of the electromagnetic spectrum. Depending on the nature of the optical companion, 
the mass transfer can take place differently between two sub-classes of HMXBs viz. super-giant X-ray 
binaries and Be/X-ray binaries. The compact object in super-giant X-ray binaries (SGXBs) captures 
sporadically a small fraction of the stellar wind from its super-giant companion which loses mass at 
a rate of 10$^{-5}$ -- 10$^{-7}$~\msol~yr$^{-1}$ \citep{Mart2017}. The luminosity of the compact object 
reaches as high as 10$^{35}$ erg~s$^{-1}$ in most of SGXBs. Luminosity with one to three orders of 
magnitude higher has also been observed from a couple of SGXB systems such as Cen~X-3, SMC~X-1, and 
LMC~X-4 where a disk-fed accretion via the Roche-lobe overflow occurs (see, e.g., 
\citealt{Reig2011, Walter2015}).

Be/X-ray binary (BeXB) systems represent a majority of the HMXB population. It consists of a neutron 
star and a non-supergiant donor of OB spectral class that shows hydrogen and helium emission lines at 
a certain point of its evolution  (\citealt{Reig2011} for a review). These emission lines are known to 
originate from an equatorial circumstellar disk formed around the Be star \citep{Porter2003}. The 
neutron star in these systems usually revolves in a wide, eccentric orbit and is known to accrete 
matter from the circumstellar disk. The abrupt mass accretion by the neutron star, at the closest 
approach (periastron), leads to the occurrence of intense X-ray outbursts with luminosity in the 
range of 10$^{36}$--10$^{38}$ erg~s$^{-1}$ \citep{Naik2013, Reig2013, Wilson2018, Jaisawal2019}. 

Strong magnetic field of the neutron stars, with a field strength of B$\sim$10$^{12}$~G in HMXBs, 
guides the accreted matter beyond the magnetospheric radius \citep{Revnivtsev2015}. The field lines 
funnel the plasma onto a confined region at the magnetic poles forming hot spots and accretion columns  
on the neutron star surface \citep{Basko1975, Nagase1989}. X-ray pulsations from the neutron star are 
observed when the hot spot rotates around the spin axis of the system. Pulse profiles of these pulsating
neutron stars (pulsars) provide information on emission geometry and also on the distribution of matter 
in its close proximity. A typical energy spectrum of a pulsar is shaped by thermal and bulk Comptonization 
processes in the accretion column \citep{Becker2007, Farinelli2012} that can be described by an empirical 
power law model modified with a high energy cutoff (see, e.g., \citealt{White1983, Paul2011, Caballero2012}).

\source (X~1908+075) is a Galactic X-ray source that was discovered by {\em Uhuru} in the 
early seventies  \citep{Giacconi1974, Forman1978, Wen2000}. The compact object in this system accretes 
from an OB supergiant companion in a 4.4 days orbit \citep{Wen2000, Levine2004, Morel2005}. 605~s pulsations 
detected from the \rxte observations, established the source as an accreting X-ray pulsar \citep{Levine2004}. 
Using the Doppler delay curves, the orbital parameters of the system such as the inclination 
(38\degree--72\degree), orbital separation (60--80 lt-s), epoch for mean longitude of 90 degrees 
$\tau_{90}$ (52631.383 MJD) along with the mass of the companion (9--31 \msol) and radius 
($\le$22 \rsol) were derived by \citet{Levine2004}. The optical companion was first thought 
to be a Wolf-Rayet star based on the X-ray analysis \citep{Levine2004}. Near-infrared observations 
of the field within the error box of HEAO-1/A3 discovered a late O-type supergiant star 
\citep{Morel2005}. Recently, \citet{Martnez2015} studied the H- and K-band infrared spectra 
of the optical companion using a non-LTE stellar atmosphere code and found the donor properties 
are mainly consistent with an early B-type (B0--B3) star. These authors also estimated the 
following donor parameters such as the stellar mass 15$\pm$6~\msol, radius 16~\rsol,  the 
effective temperature T$_*$ = 23.0$^{+6}_{-3}\times10^3$~K, luminosity log(L/\lsol) = 4.81$\pm$0.25, 
and the terminal wind velocity $v_\infty$ = 500$\pm$100~\kms. The source distance is known to be
4.85$\pm$0.5~kpc depending on the extinction curve \citep{Martnez2015}.

Long term timing studies showed that the spin period of the pulsar changes erratically on years 
time scale. It is only possible when mass accretion through the capture of stellar wind occurs in 
the system \citep{Furst2011, Furst2012}. A similar conclusion was drawn based on the detection of 
the Compton shoulder in the iron fluorescence line. Such a line feature appears when the Compton 
thick medium surrounds the accreting object \citep{Torrejon2010}. The pulse profile of the pulsar 
is known to be strongly energy dependent. It evolves from a broad structure in soft X-rays to a 
narrow peak at higher energies \citep{Furst2011, Furst2012, Jaisawal2013}. Broad-band energy 
spectrum of the pulsar is consistent with an absorbed high energy cutoff power law model along 
with a blackbody component \citep{Furst2011, Furst2012, Jaisawal2013}. Based on a 
possible/tentative detection of cyclotron absorption line at $\sim$44 keV \citep{Jaisawal2013}, 
the magnetic field of the neutron star is estimated to be $\sim$3.8$\times$10$^{12}$~G. 
Superorbital modulation at a period of 15.2 days is also known from the system \citep{Corbet2013}.   

In the present paper, we study the pulse profile and spectral evolution of \source using  
\nustar and \astrosat observations in 2015 and 2017 July, respectively. We  examine 
the distribution of stellar wind during these monitorings. 
The spin period evolution of the pulsar is also studied based on the long term measurements. 
The details of the observations, results, discussion, and conclusion are summarized in 
Section~2, 3, and 4, respectively.

\begin{table}
\centering
\caption{Log of 4U~1909+07 observations with {\it NuSTAR}, {\it Swift}, and {\it Astrosat}.}
\begin{tabular}{ccccc}
\hline
\hline

Observatory/	   &ObsID       &Start Date     &Expo.  \\
Instrument	       &	        &(MJD)				  &(ks) \\
\hline

{\it NuSTAR}  	 &30101050002  &57204.65     	&41.3 \\
{\it Swift}/XRT &00081661001  &	57205.38 	&1.7 \\
{\it Astrosat}   &A03\_114T01\_9000001390   &57951.27    	&30.3 \\ 

 \hline
\hline
\end{tabular}
\label{log}
\end{table}

\begin{figure*}
 \begin{center}$
 \begin{array}{ccc}
 \includegraphics[height=3.35in, width=2.65in, angle=-90]{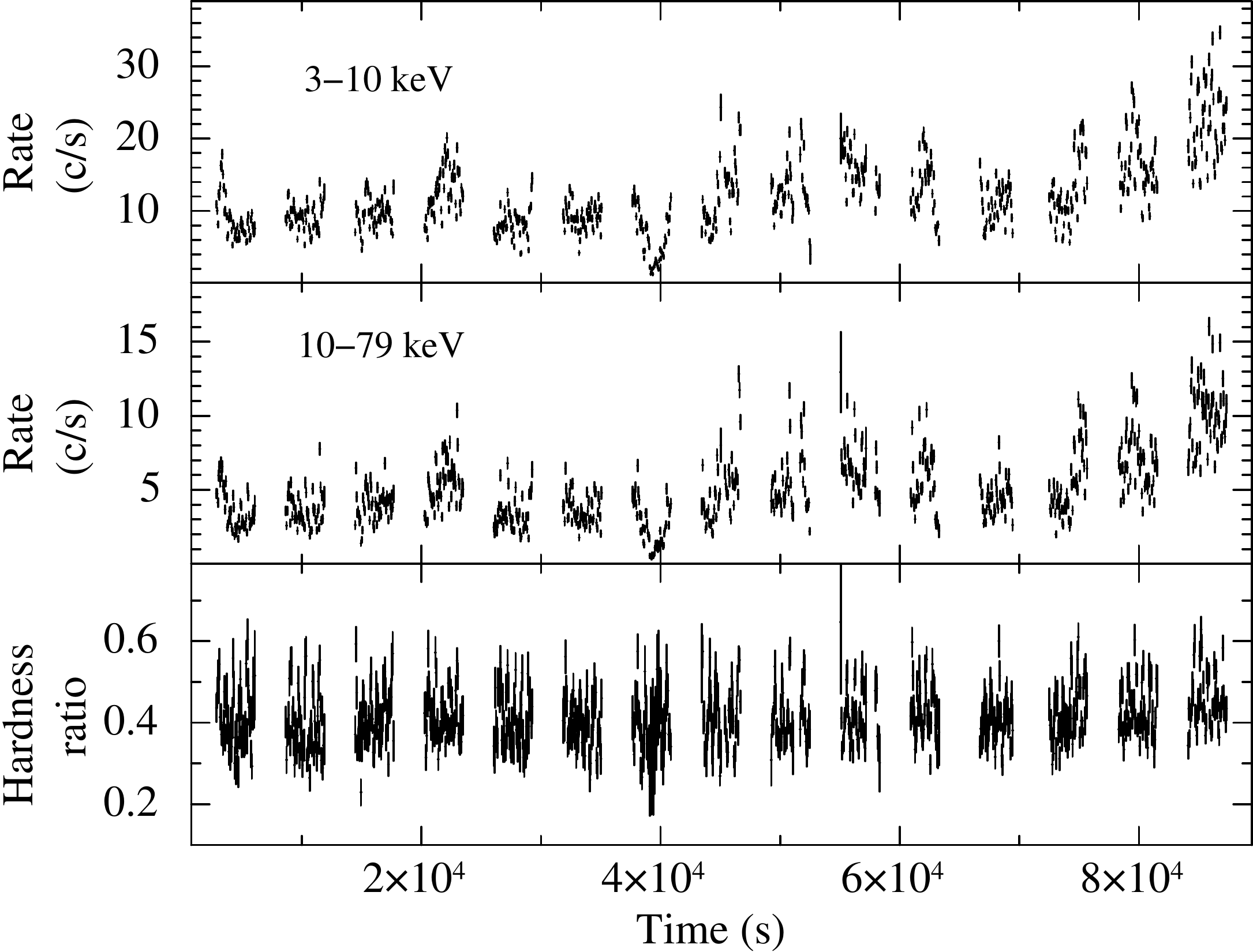} &
 \includegraphics[height=3.35in, width=2.65in, angle=-90]{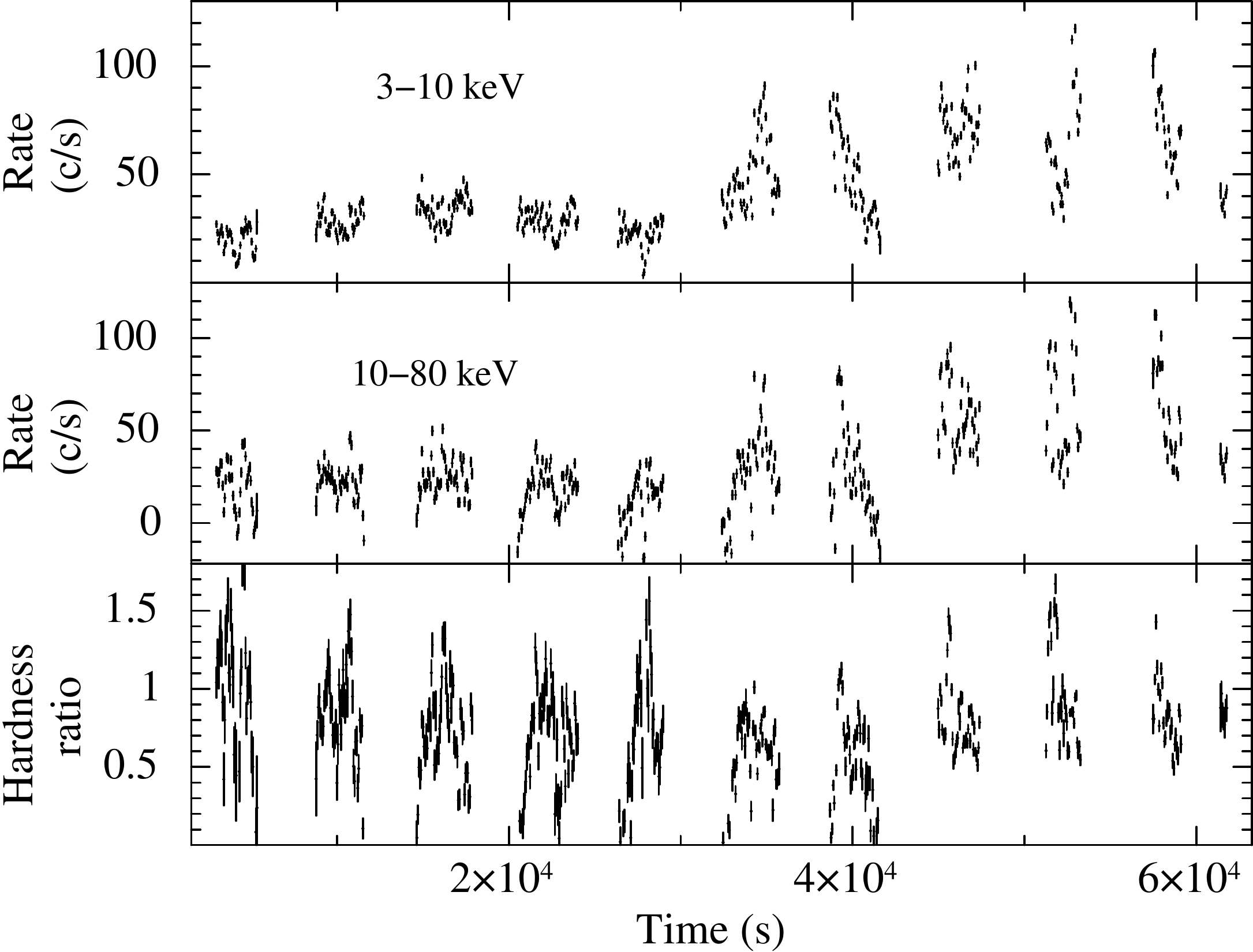} & 
 \end{array}$
 \end{center}
\caption{Soft and hard X-ray light curves of the pulsar 4U~1909+07 obtained from the \nustar 
(left) and \astrosat/LAXPC (right) observations at one tenth time resolution of the spin period 
are shown in top two panels of the figure.  Several flare-like features are visible in the 
light curve. A ratio between the hard and soft X-ray light curves i.e. hardness ratio is also 
presented in respective bottom panels.}
\label{fig-lc}
\end{figure*}  


\begin{figure}
\centering
\includegraphics[height=2.5in, width=3.3in, angle=0]{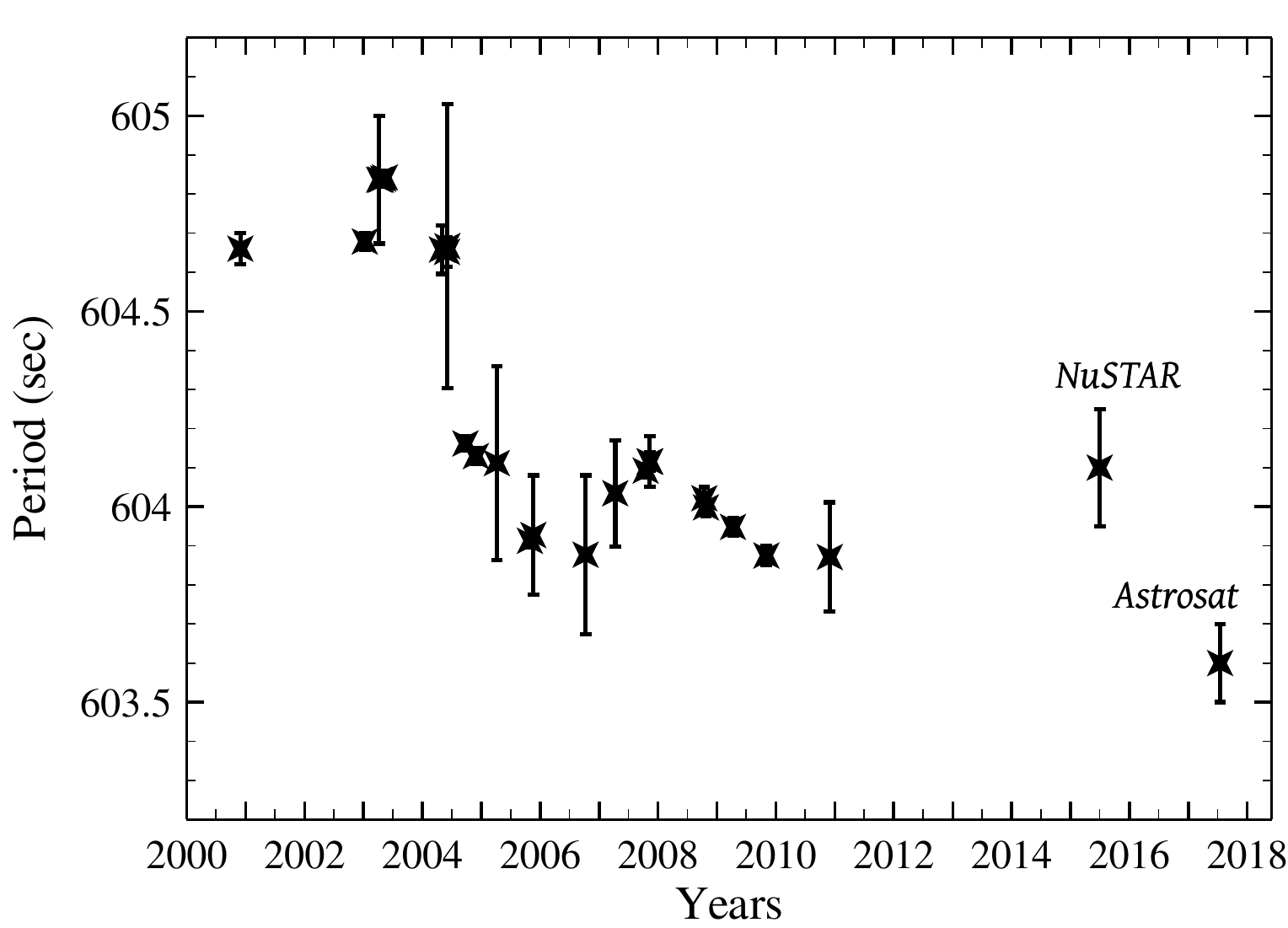}
\caption{Long term spin period evolution of \source based on measurements provided 
by \citet{Levine2004}, \citet{Furst2011}, \citet{Furst2012}, and \citet{Jaisawal2013} 
using \rxte, {\em INTEGRAL}, and \suzaku data. The spin period measurements 
from the recent \nustar and \astrosat observations are also included in the figure.}. 
\label{fig-spin}
\end{figure}


\begin{figure}
\centering
\includegraphics[height=3.2in, width=2.85in, angle=-90]{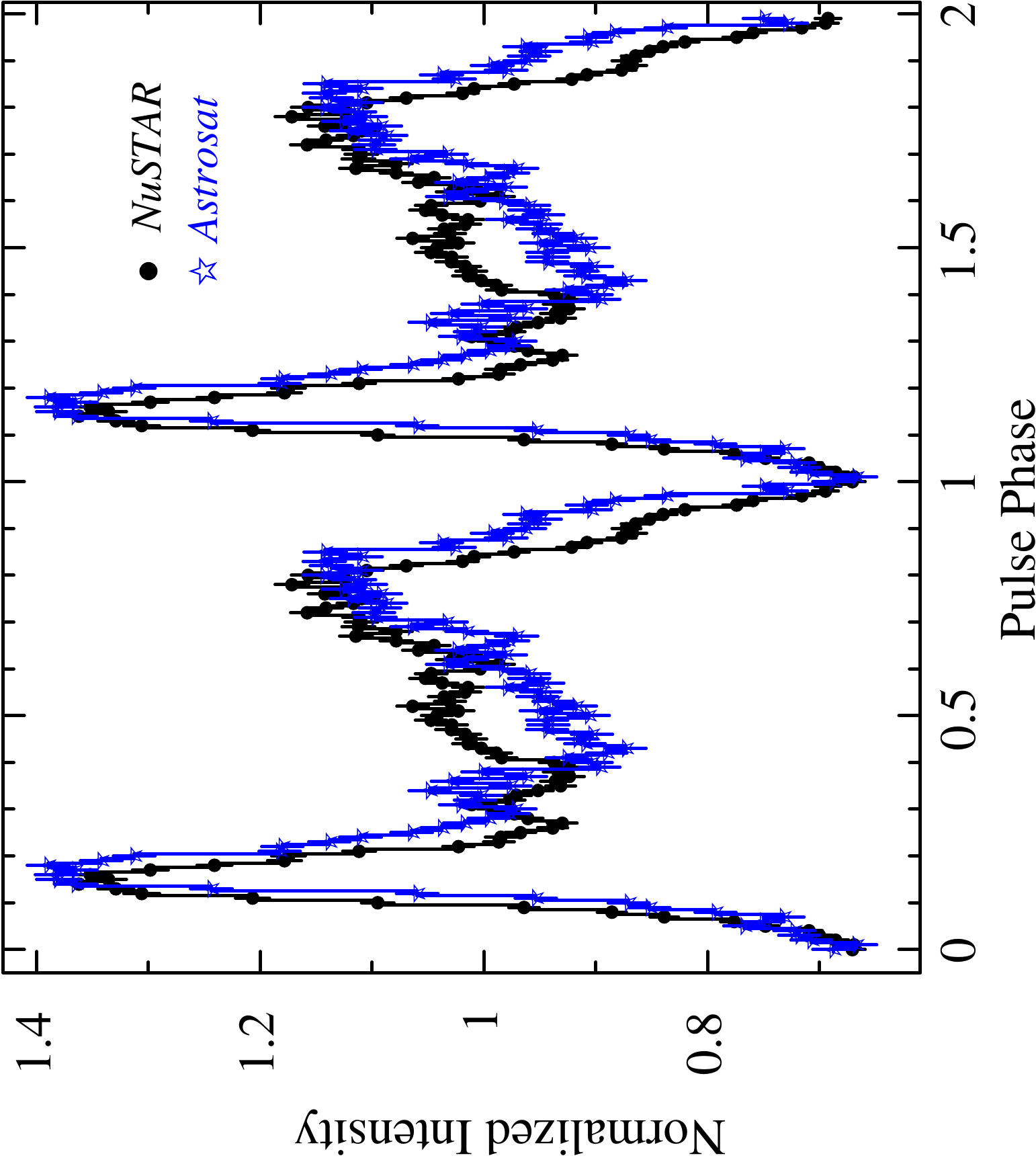}
\caption{Pulse profiles of \source obtained from the 2015 July and 2017 July observations with 
\nustar and \astrosat in a 3-79~keV range, respectively. The structure of the pulse profiles 
from both the instruments in broad energy range is found to be almost identical except a major 
difference at around 0.5 phase. Pulse profiles are normalized with respect to average intensity.
The error bars represent 1$\sigma$ uncertainties. Two pulses are shown for clarity. The pulse profiles from \nustar and \astrosat data are obtained by considering 100 phase bins per period. } 
\label{pp}
\end{figure}

\begin{figure*}
 \begin{center}$
 \begin{array}{ccc}
 \includegraphics[height=3.35in, width=3.2in, angle=-90]{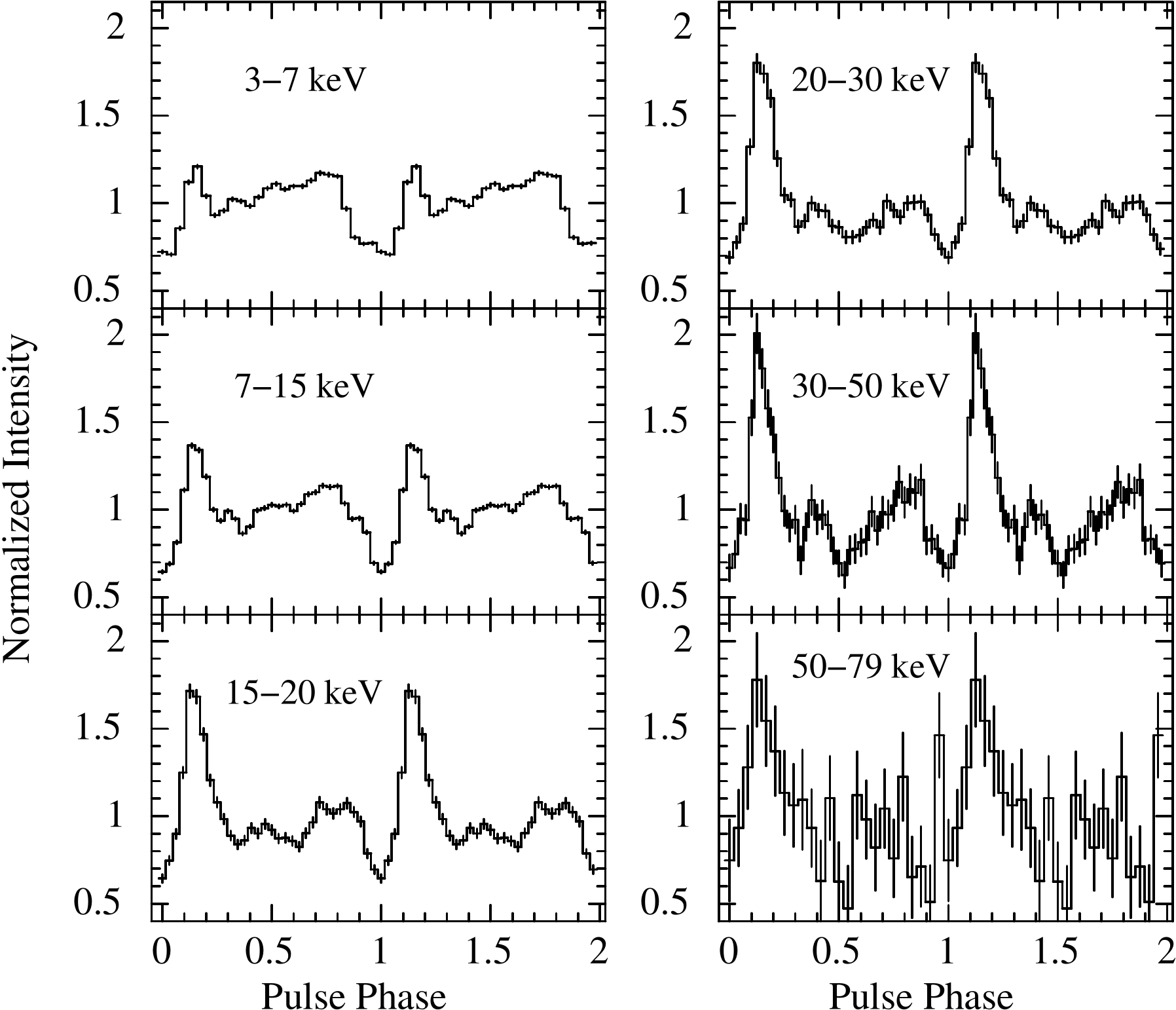} &
 \includegraphics[height=3.35in, width=3.2in, angle=-90]{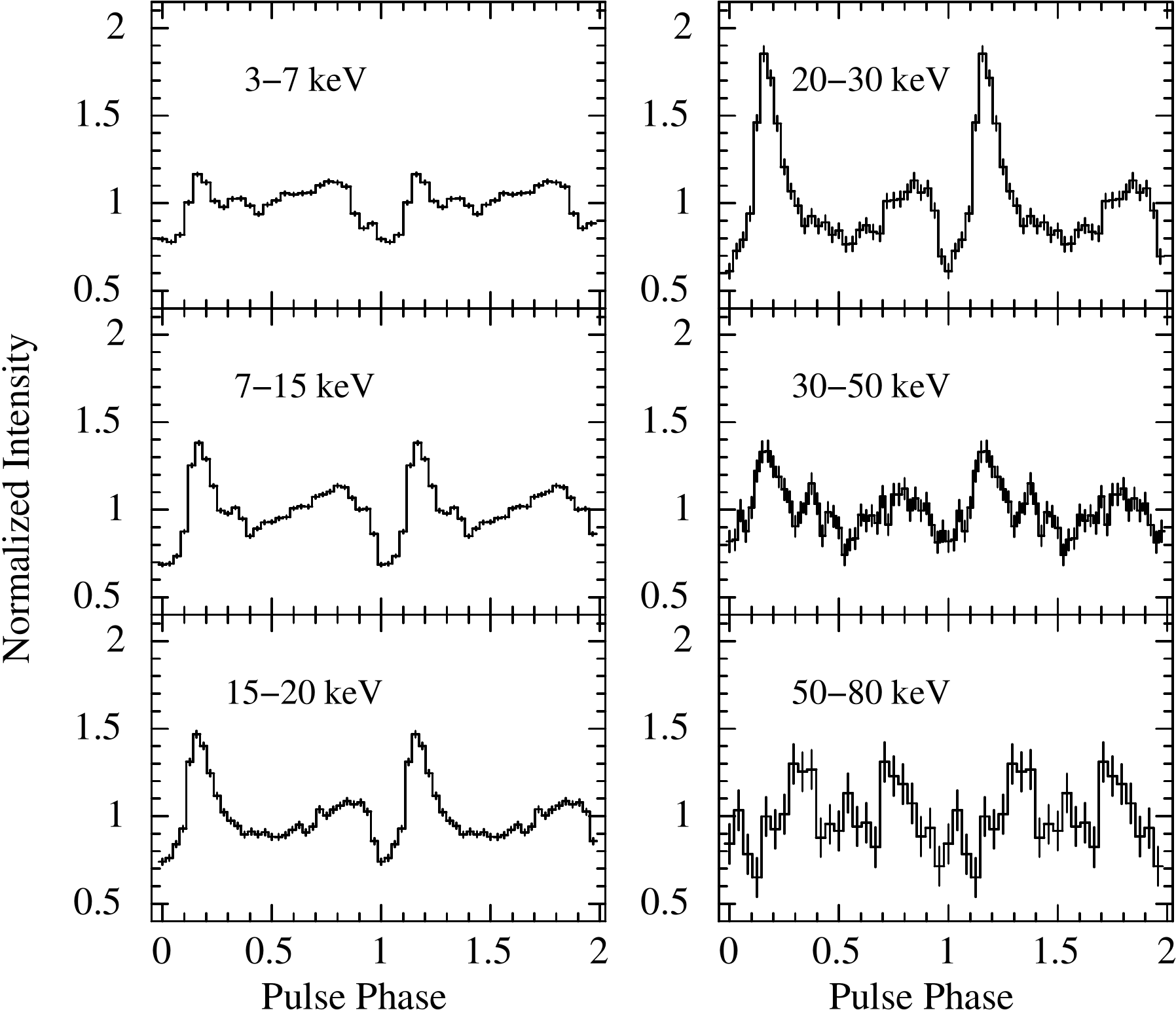} & 
 \end{array}$
 \end{center}
\caption{Energy resolved pulse profiles of \source obtained by folding the light curves 
from \nustar (left) and \astrosat/LAXPC (right) observations with the estimated spin 
periods. Two pulses are shown in each panel for clarity. The error-bars represent 1$\sigma$ 
uncertainties.}
\label{profile}
\end{figure*}


\begin{figure}
\centering
\includegraphics[height=3.in, width=3.in, angle=0]{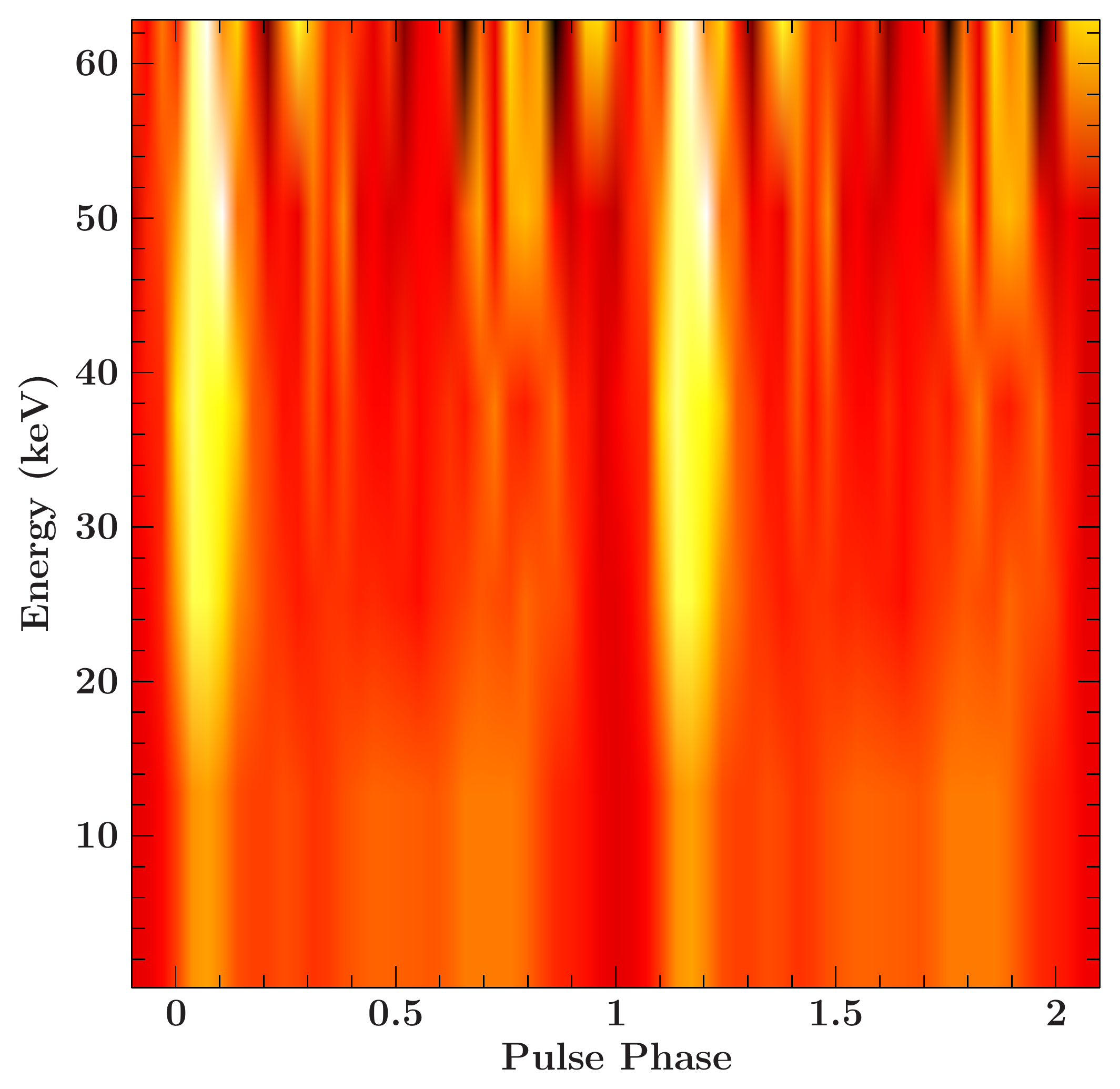}
\caption{Color-coded pulse profile map produced from the energy resolved
light curves of the pulsar using \nustar data in 2015 July. The color in the figure 
represents the normalized intensity of the profile, varying from red (low) to white (high). 
Two pulses are shown for clarity.}
\label{map}
\end{figure}

\section{Observations and Data Analysis}
\label{sec:data}

\subsection{\nustar and \swift}

\nustar is a first hard X-ray focusing observatory that was launched by NASA in 2012 June 
\citep{Harrison2013}. It is sensitive in 3--79 keV energy range and reflects the photons at 
grazing angle incidence. \nustar consists of two multi-layer coated focal plane mirror modules, 
named as FPMA and FPMB. \source was observed with \nustar on 2015 July 01 for exposure of 
41.3 ks (Table~\ref{log}). Standard procedures were followed to analyze the raw data with the 
help of {\tt NuSTARDAS} 1.6.0 software in HEASoft version 6.24. We reprocessed unfiltered events 
from the FPMA and FPMB by using {\it nupipeline} routine and updated CALDB of version 20191219. 
We then extracted source products by selecting a circular region of 150~arcsec radius with the 
source co-ordinates as the center, in the {\tt DS9} image of cleaned events, using the {\it 
nuproducts} task. The background light curves and spectra were accumulated in a similar manner 
by selecting a source-free circular region of 150~arcsec radius. \source was also monitored by 
\swift/XRT on 2015 July 02 for an effective exposure of 1.7 ks in photon counting mode. We 
considered \swift observation for contemporary spectral analysis with \nustar. The XRT data are 
processed by using the online standard tools provided by the UK Swift Science Data 
Centre\footnote{\url{http://www.swift.ac.uk/user_objects/}} \citep{Evans2009}. 
 For timing analysis, the light curves from FPMA and FPMB units 
of \nustar were combined in our study.

\subsection{\astrosat}

\astrosat is the first Indian astronomical mission launched by ISRO in September 2015 \citep{Agrawal2006}. 
It is capable of observing stellar sources in multi-wavelength bands by using five sets of instruments 
such as Soft X-ray Telescope (SXT; \citealt{Singh2017}), Large Area X-ray Proportional Counters (LAXPCs; 
\citealt{Agrawal2017,Antia2017}), Cadmium Zinc Telluride Imager (CZTI; \citealt{Rao2017}), a Scanning 
Sky Monitor (SSM; \citealt{Ramadevi2018}), and Ultraviolet Imaging Telescope (UVIT; \citealt{Tandon2017}).
\astrosat observed \source on 2017 July 17 for an effective exposure of 30.3 ks.  In the present study, 
we used data from the SXT and LAXPC instruments as the source was very faint for the CZTI. The UVIT was 
not operational during the observation. The SXT is a soft X-ray focusing telescope consisting of a CCD 
detector sensitive in a 0.3-8 keV energy range. We reprocessed the SXT data by using standard pipeline and 
merging tool {\tt sxtevtmergertool} provided by \astrosat Science Support Cell 
(ASSC\footnote{\url{http://astrosat-ssc.iucaa.in/}}). The source spectrum was extracted thereafter
using {\tt XSELECT} package by considering a circular region of 4~arcmin radius with source coordinates 
as centre, on the SXT CCD chip. The background spectrum was obtained from a blank sky observation. The 
LAXPC is the primary X-ray instrument onboard \astrosat which is sensitive to photons in a 3-80 keV energy 
range. It consists of three identical units providing a total effective area of 6000 cm$^2$ at 15 keV. 
In our study, the raw event analysis mode data are reprocessed through the standard routines available 
in {\tt LAXPCsoftware}. We used this package to create source light curves and spectral products. The 
same observation was considered for the LAXPC background during the Earth occultation period. For timing 
studies, we have combined the light curves from LAXPC10 \& 20 in our analysis. The data from LAXPC30 are 
not used due to drastic change in the instrument gain.

\section{Results}

\subsection{Timing analysis}

Using the orbital ephemeris of the binary system as reported by \citet{Levine2004}, 
we found that the pulsar was observed in 0.215--0.437 and 0.87--0.05 phase ranges 
of  the 4.4 days binary orbit with \nustar and \astrosat observatories, respectively.
Background subtracted light curves obtained from these observations are shown in 
Figure~\ref{fig-lc}.  Flare-like features can be seen in the soft and hard X-ray
light curves (top two panels in each side of the figure) as generally seen in the light
curves of other SGXBs due to inhomogeneous mass accretion  
from the stellar wind of the companion star (see, e.g., \citealt{Naik2011, Odaka2013}). Bottom 
panel in each side of the figure represents the hardness ratio (HR) which is defined 
as the ratio between the hard X-ray  (10-79 keV range for \nustar data and 
10-80 keV range for \astrosat/LAXPC data) and soft X-ray  (3-10 keV range) 
light curves. Though we observe several flare-like features in the soft and hard 
X-ray light curves, the HRs remain almost constant throughout these monitoring. This 
suggests that the spectral shape of the pulsar was unchanged during the \nustar and 
\astrosat observations. 

We searched for pulsation in the X-ray light curves by using 
the $\chi^2$-maximization technique \citep{Leahy1987}. The barycentric corrected pulse 
period of the neutron star was estimated to be  604.1$\pm$0.15~s and 603.6$\pm$0.1~s 
from the \nustar and \astrosat light curves binned at a time resolution of 0.1 s, respectively. 
The errors in the pulse period are estimated for a 1$\sigma$ confidence interval.

Pulsation periods and the associated uncertainties were also calculated by fitting the pulse times of arrival (TOA) with a timing model.
We followed the methodology described by \citet{Ray2011}, that has been applied to other  HMXB pulsars \citep{2019ApJ...879..130R,2019MNRAS.488.5225V,2020MNRAS.494.5350V}. 
We subdivided the data into 10-20 intervals and generated TOAs by comparing the pulse profile of each interval with a template profile.
We fitted the TOAs to a timing model with constant period.
The best pulse period is consistent with the main peak found with the epoch folding method, while uncertainties were found to be of the order of 0.06-0.1 s. Nevertheless, the solution has some timing residuals likely caused by changes of the pulse profile within each observation.

Long term spin 
period evolution of \source is given in Figure~\ref{fig-spin}. The spin periods of the pulsar 
at earlier epochs were taken from \citet{Levine2004}, \citet{Furst2011}, \citet{Furst2012}, 
and \citet{Jaisawal2013} using \rxte, {\em INTEGRAL}, and \suzaku data. We also
included our pulse period measurements in the figure that clearly signifies a global 
spin-up trend of the pulsar since its discovery,  though a sharp drop in the period
has been observed during 2004--2005. The pulse period is effectively changed from 
604.7 s to 603.6 s between 2001--2017 resulting an average spin-up rate $\dot{P}$ of 
1.71$\times$10$^{-9}$~s~s$^{-1}$. 

Following the estimation of spin period of \source, the light curves from \nustar (FPMA+FPMB) and added light curves of LAXPC10 and LAXPC20 of \astrosat are folded with respective periods to obtain corresponding pulse profiles.  The pulse profiles of the pulsar with 100 phase 
bins per period, obtained from both the data sets are shown in Figure~\ref{pp}. The observed pulse morphology 
shows a narrow emission peak at around 0.1 phase, followed by a broad structure at other pulse 
phases. The emission geometry appears to be similar during the 2015 and 2017 observations, though 
marginal differences can be spotted at certain pulse phases,  especially at $\sim$0.5 phase, 
see Figure~\ref{pp}. To understand the energy evolution of the emission geometry, the energy
resolved pulse profiles are generated and shown in Figure~\ref{profile}. Using \nustar data, we 
observed a broad profile below 7 keV that evolves into a narrow emission peak, followed by a 
plateau like structure at higher energies (left side of Figure~\ref{profile}). The color-coded 
pulse profile map shows this variation clearly (Figure~\ref{map}). This map is generated by 
using \texttt{ISISscripts}\footnote{\url{http://www.sternwarte.uni-erlangen.de/isis/}}. 
The color in the map represents the normalized intensity of the pulse profiles ranging 
from red (low) to white (high). A similar kind of pulse profile evolution is also detected 
during the \astrosat observation  with a clear difference above 30 keV (right side of
Figure~\ref{profile}). This difference in the hard X-ray pulse profiles possibly arises due 
to the faintness of the source at hard X-rays, combined with the presence of a systematic 
uncertainty in the LAXPC background measurement beyond $\sim$30 keV (refer to 
\citealt{Antia2017, Misra2017} for information on the LAXPC background).

\begin{figure}
\centering
\includegraphics[height=3.3in, width=2.4in, angle=-90]{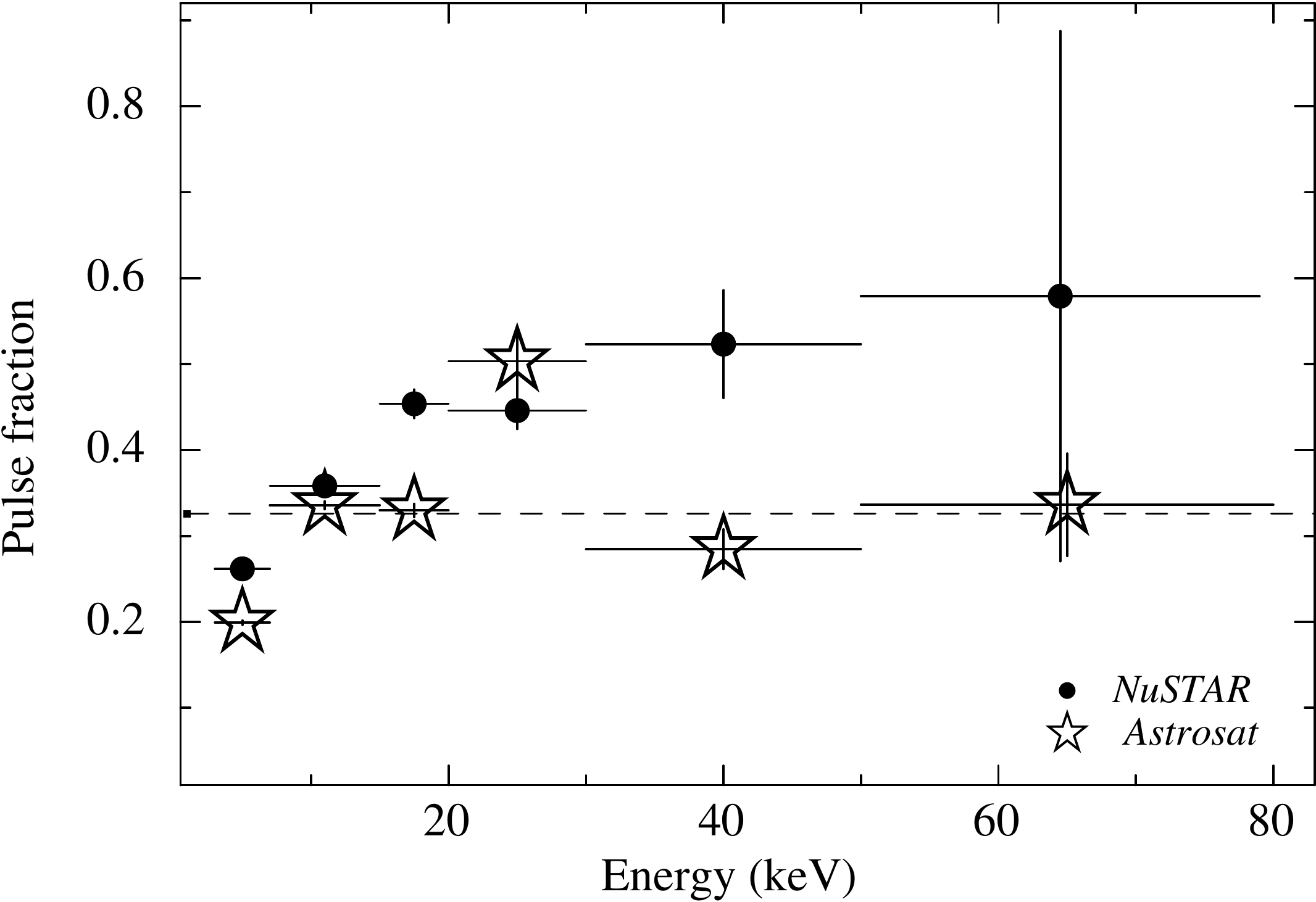}
\caption{Pulse fraction variation of the pulsar with energy obtained 
from the light curves in multiple energy bands. The dashed line represents 
an averaged value of pulse fraction $\approx$32\% estimated from  \nustar 
and \astrosat data in 3-80 keV band.}
\label{pf}
\end{figure}

We define the pulse fraction (PF) of the pulsar in the following manner -
\begin{equation} \label{eq1}
PF = \frac{F_{max} - F_{min}}{F_{max} + F_{min}},
\end{equation}
where {\em F$_{max}$} and {\em F$_{min}$} correspond to maximum and minimum intensities observed 
in the profile, respectively. During both the observations, the average values of the PF 
are found to be $\approx$32\%. To evaluate the changes in PF with energy, we investigated the 
energy resolved profiles and estimated corresponding pulse fraction which are shown in 
Figure~\ref{pf}.  It can be seen that the PF values increase with energy as seen in 
several other accretion powered X-ray pulsars \citep{Lutovinov2009, Jaisawal+18}. The PF of
the pulsar \source, estimated from the \nustar and \astrosat observations are comparable up to 
30 keV, beyond which the PF values from \astrosat observation decreased significantly
(Figure~\ref{pf}). This is possibly due to the presence of systematics in the LAXPC background 
measurements at high energies (as mentioned earlier).

\begin{figure}
\centering
\includegraphics[height=3.3in, width=2.9in, angle=-90]{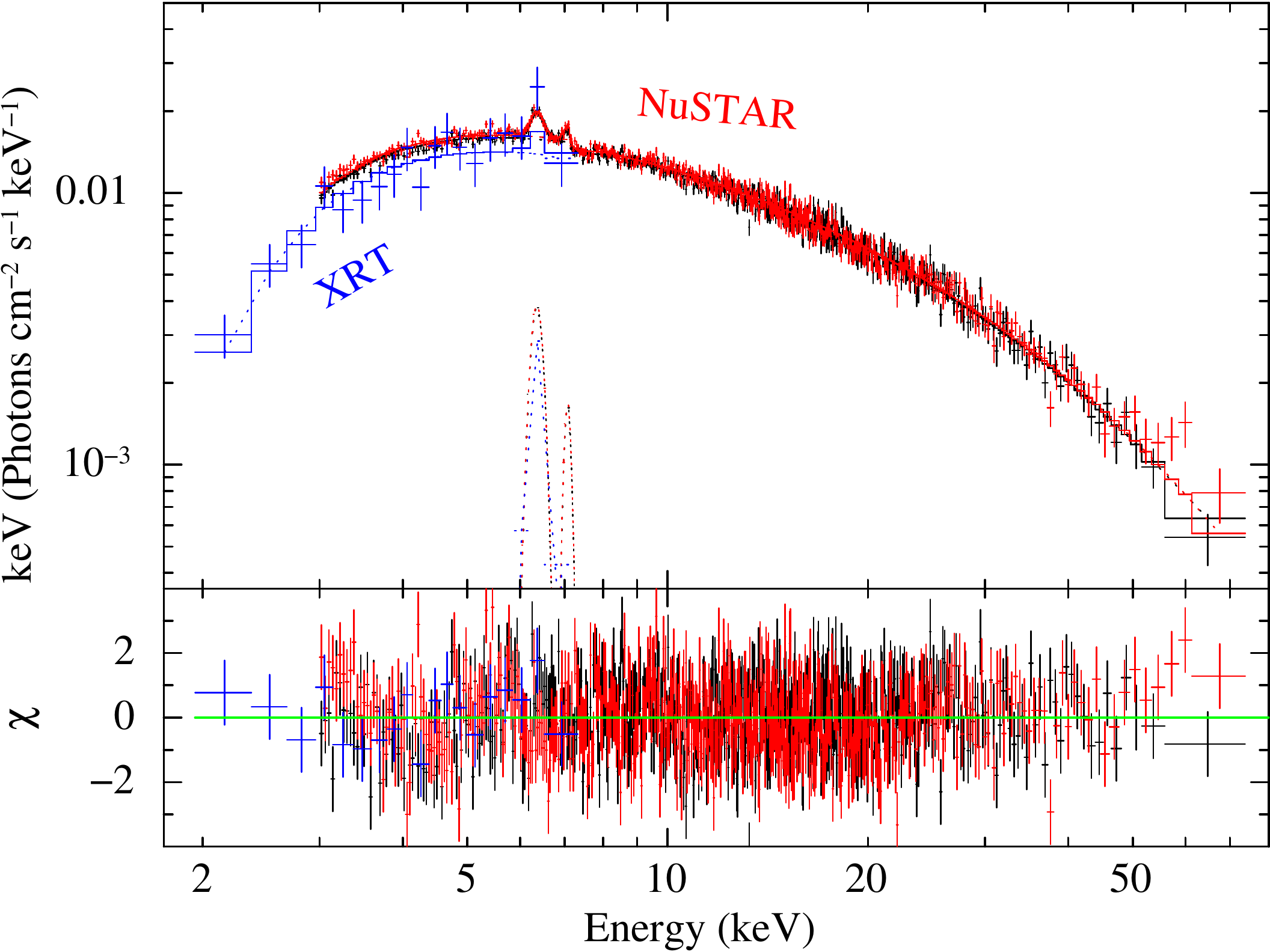}
\caption{The 1-70 keV energy spectrum of the pulsar obtained from the {\it Swift}/XRT and \nustar 
observations in 2015 July. Best fitted energy spectrum and its corresponding spectral residuals 
are shown in top and bottom panels, respectively. }
\label{spec_nu}
\end{figure}
\begin{figure}
\centering
\includegraphics[height=3.3in, width=2.9in, angle=-90]{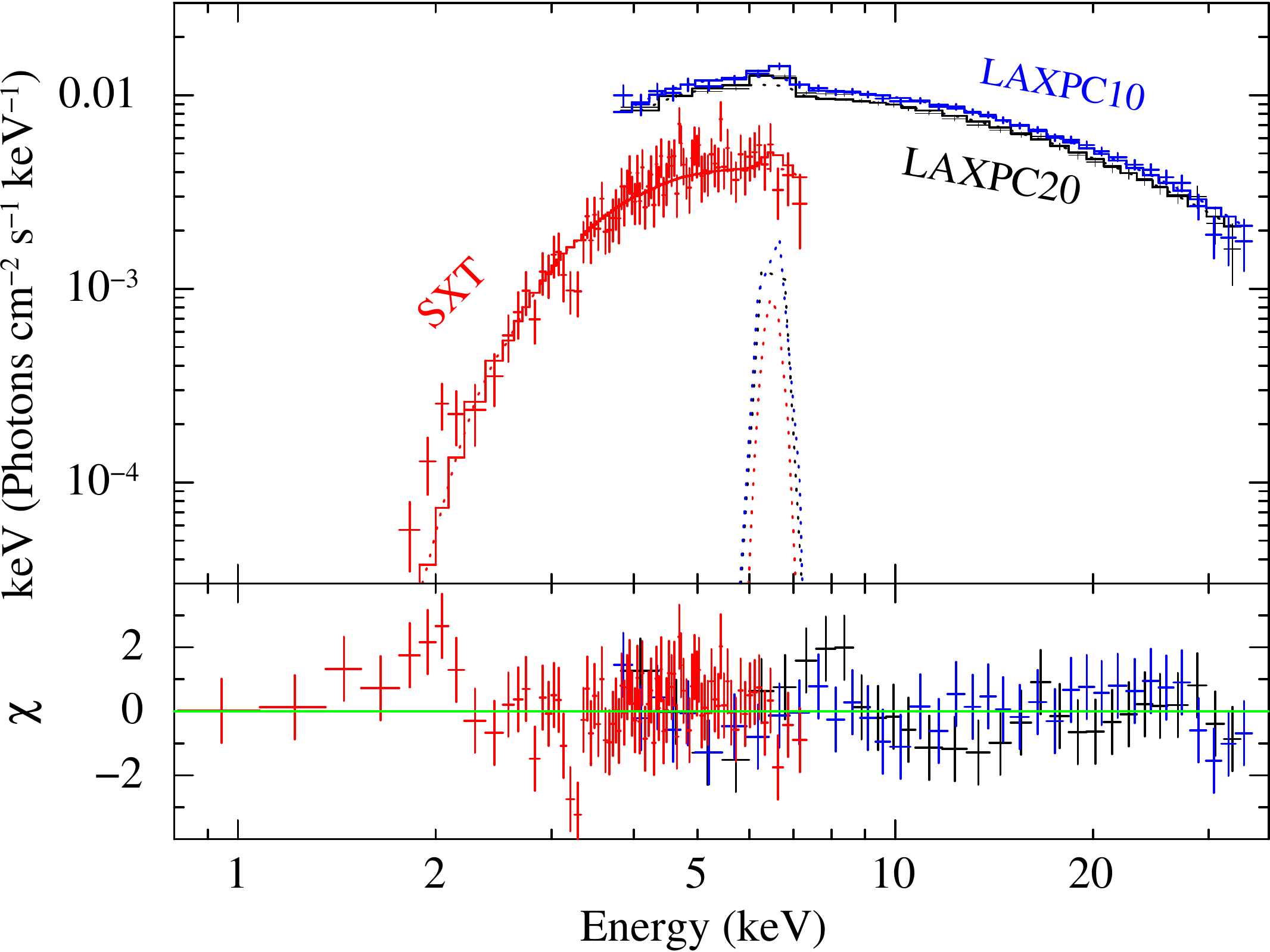}
\caption{The 1-35 keV energy spectrum of the pulsar obtained from the SXT and LAXPC instruments 
on-board \astrosat in 2017 July. Best fitted energy spectrum and corresponding spectral residuals 
are shown in top and bottom panels, respectively.}
\label{spec_ast}
\end{figure}

\subsection{Pulse-phase-averaged spectroscopy}

The energy spectrum of \source is investigated to understand its emission characteristics during 
the \nustar, \swift, and \astrosat observations. Standard empirical models were chosen to fit 
the broadband spectrum of the pulsar in {\tt XSPEC} package \citep{Arnaud1996}. The 1-79 keV 
spectrum of the pulsar from \swift/XRT and \nustar is considered in our analysis. We also fitted 
the 1-35~keV energy continuum by using the data from SXT and LAXPC10 \& 20 instruments on-board 
\astrosat. We found that an absorbed high energy cutoff power law 
model\footnote{\url{https://heasarc.gsfc.nasa.gov/xanadu/xspec/manual/Models.html}} describes 
these spectra well with the goodness of fit per degree of freedom $\chi^2_\nu$=$\chi^2/\nu$ $\approx$1 in 
each case (Figures~\ref{spec_nu} \& ~\ref{spec_ast}). In the fitting, the contribution for 
local and interstellar medium absorption $N_{\rm H}$ observed along the source direction is 
expressed by {\tt TBabs} component \citep{Wilms2000}. We also detected iron fluorescence lines 
at 6.4 and 7.1~keV in the \nustar spectrum of the source. Any signature of cyclotron absorption 
line is not found in these observations. Best fitting spectral parameters are given in 
Table~\ref{table2}. Except for the absorption column density, other components like photon index, 
folding and cutoff energies are consistent (within the errors) between the 2015 and 2017 data. 
\source is found to be strongly obscured owing to a larger column density $N_{\rm H}$, in 
addition to the interstellar absorption of 1.21$\times$10$^{22}$~cm$^{-2}$ along the source 
direction \citep{HI4PI2016}. We estimated the total flux using $\tt cflux$ convolution model 
in this paper. For a source distance of 4.85~kpc, the 1-80 keV luminosity of the pulsar is 
measured to be 1.67 and 1.24$\times$10$^{36}$ erg~s$^{-1}$ during 2015 and 2017 observations, 
respectively.

\begin{table}
\centering
\caption{Best-fitting spectral parameters (with 90\% errors) of \source obtained from 
the {\it NuSTAR} and \swift/XRT observations in  2015 July, and  \astrosat observation 
in 2017 July. The fitted model consists of an absorbed high-energy cutoff power 
law (HECut) with Gaussian components for iron emission lines.}
\begin{tabular}{ |l | cc}
\hline 
Parameters                      &  \multicolumn{1}{c}{\nustar+\swift}  &\multicolumn{1}{c}{\astrosat}   \\ 
&  \\\cline{1-3}  \\
                                &HECut     &HECut     	 \\
\hline
N$_{\rm H}$$^a$                    &10.1$\pm$0.5         &18$\pm$3     \\
Photon index                       &1.52$\pm$0.04        &1.5$\pm$0.3     \\
E$_{\rm cut}$ (keV)	               &8.0$\pm$0.3          &7$\pm$3 \\
E$_{\rm fold}$ (keV)	             &26$\pm$1             &23$\pm$8 		\\

\\
{\it Emission lines } \\
Line energy (keV)              &6.36$\pm$0.03       &6.5     \\
Eq. width  (eV)                &90$\pm$14	         &154$\pm$66     \\
Line energy (keV)              &7.1$\pm$0.1        &--     \\
Eq. width (eV)                 &27$\pm$10	         &--        \\

\\
{\it Source flux}$^b$
 \\
Flux (1-30 keV)   		        &4.93$\pm$0.05      &3.7$\pm$0.2    \\
Flux (1-80 keV)    		        &5.92$\pm$0.06      &4.4$\pm$0.3      \\ 
\\
$\chi^2_\nu$ ($\nu$)      &1.12 (1001)          &0.91 (143)        \\
\hline
\end{tabular}
\flushleft
$^a$ : Equivalent hydrogen column density in 10$^{22}$ cm$^{-2}$,
$^b$ : Unabsorbed flux in the unit of 10$^{-10}$ ~erg~cm$^{-2}$~s$^{-1}$. 
\label{table2}
\end{table}

\begin{figure}
\centering
\includegraphics[height=3.3in, width=3.1in, angle=-90]{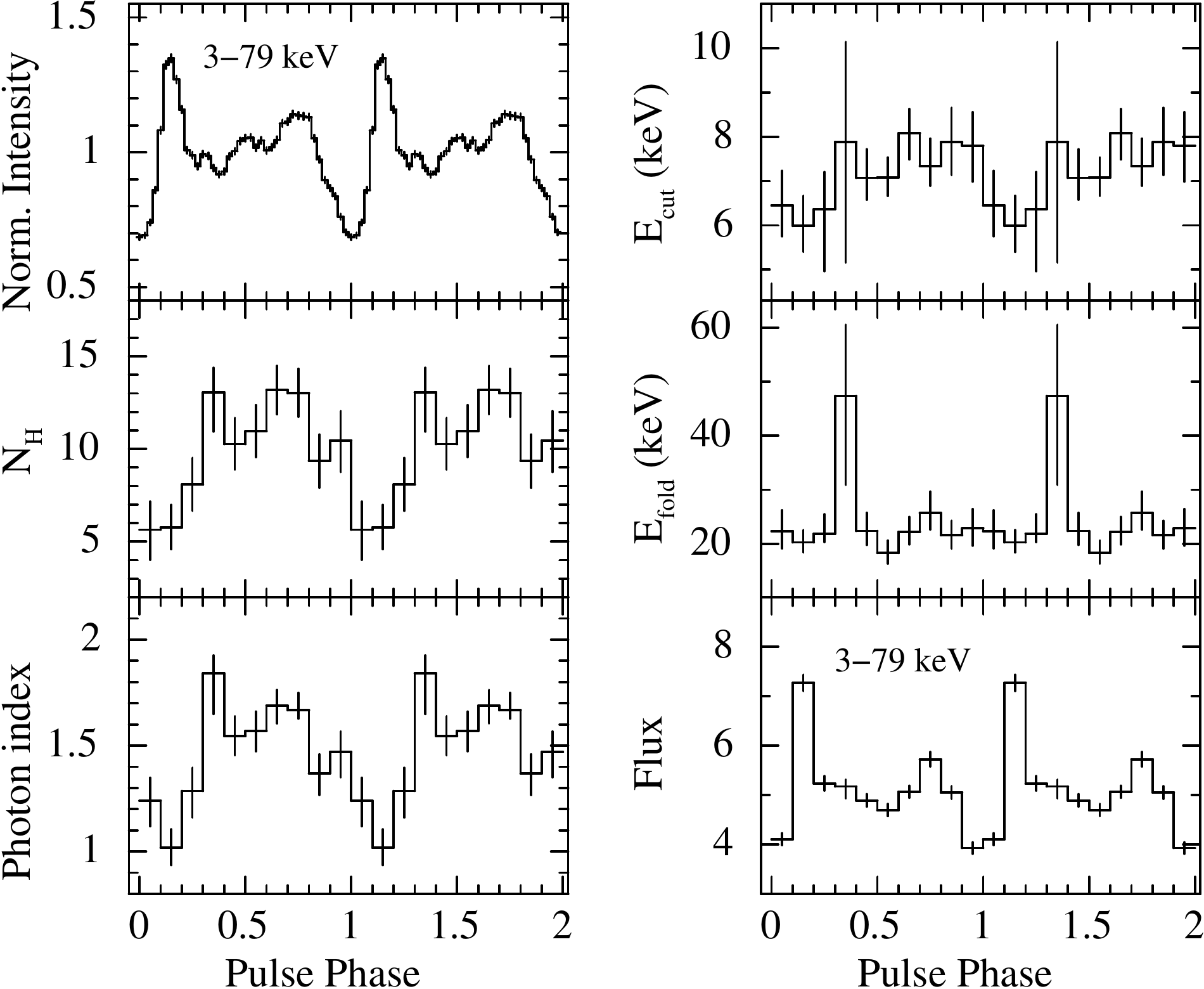}
\caption{Phase-resolved spectroscopy of \source using {\em NuSTAR} data in 
2015 July. The first, second, and third panels on left side of the figure
show the 3-79 keV pulse profile, absorption column density (10$^{22}$~cm$^{-2}$), 
and power-law photon index, respectively. The cutoff energy, folding energy, and the 
3-79 keV unabsorbed flux (10$^{-10}$ ~erg~cm$^{-2}$~s$^{-1}$ units) are presented 
in first, second, and third panels on right side of the figure. }
\label{prs}
\end{figure}

\subsection{Pulse-phase-resolved spectroscopy}

To investigate the changes in local accreting environment of the pulsar and its effect 
on the emission geometry, phase-resolved spectroscopy is performed by using the \nustar 
data. We extracted pulse-phase resolved spectra in a total of 10 phase bins. Considering 
appropriate backgrounds, response matrices, and effective areas, the 3-79 keV spectra 
obtained from FPMA and FPMB detectors are fitted with an absorbed HECut model. In the 
fitting, we fixed the width of iron lines at the corresponding values obtained from 
phase-averaged spectroscopy (Table~\ref{table2}). We found that the above model can 
describe each spectrum well with the value of $\chi^2/\nu$ close to 1. 

Figure~\ref{prs} shows the pulse-phase evolution of spectral parameters such as absorption 
column density, photon index, folding energy, high energy cutoff, and the source flux. The 
$N_{\rm H}$ varies between 5 and 15$\times$10$^{22}$~cm$^{-2}$ and is observed to be 
significantly larger than the expected interstellar absorption (1.21$\times$10$^{22}$~cm$^{-2}$) 
along the line of sight. This high value of column density is detected across all pulse-phases 
of the pulsar, except at 0.1--0.2~phase that coincides with the narrow emission peak in the 
profile. Corresponding to the changes in the photon index, the spectrum gets softer in the plateau 
region of the pulse profile. A harder spectrum with a photon index of 1 is detected at 0.1--0.2 
pulse-phase of the pulsar. Besides a higher value of the folding energy at a single phase 
bin (Figure~\ref{prs}), the parameters such as cutoff and folding energies are almost constant 
(within the error bars) with the pulsar rotation. The change in 3-79 keV source flux with 
pulse-phase ranges is found to be consistent with the shape of the pulse profile in the same 
energy range.

\section{Discussion and Conclusions}

 We have presented the long term spin period evolution of \source using the 
measurements provided by \cite{Furst2011, Furst2012}, and \cite{Jaisawal2013} along 
with the recent \nustar and \astrosat observations in this decade. The pulse period 
of the pulsar has been found to be in globally decline since its discovery 
(Figure~\ref{fig-spin}). The long term evolution also consists of local fluctuation 
that is considered as a random walk like behavior \citep{Furst2011}. In the case of 
wind accreting sources like Vela~X-1 \citep{Deeter1989}, such behaviour signifies the absence of a 
persistent accretion disk. However in case of \source, the observed net changes in
the pulsar period from 604.7 s to 603.6 s during 2001--2017 (Figure~\ref{fig-spin}), 
the presence of transient accretion disk can not be entirely ruled out.

\subsection{Magnetic field strength of the neutron star}

The strength of magnetic field of a neutron star (pulsar) can be directly estimated through the 
detection of cyclotron resonance scattering features in the broadband spectrum. However, 
there are only about 35 X-ray pulsars in the spectra of which such features are detected  
\citep{Jaisawal2017,Staubert2019}. In the absence of cyclotron features, however, there are 
alternative methods such as based on the evolution of the spin period or by determining 
the propeller regime through which the strength of the pulsar magnetic field has been estimated 
(\citealt{Ghosh1979, Tsygankov2016}).  We can calculate the magnetic field of a pulsar based on the magnetospheric 
plasma interaction and the spin period evolution. The accreted material carries angular momentum 
that exerts a torque on the neutron star magnetosphere when the coupling happens at the 
boundary \citep{Ghosh1979}. The neutron star gets spin-up or spin-down depending on the nature of the 
net torque applied to the magnetosphere. The accretion in the binary system usually takes 
place through the formation of an accretion disk under Roche-Lobe overflow or through the 
capture of stellar wind from the companion star. In case of disk accretion, the specific angular
momentum of matter $\omega{R{_A^2}}$ exceeds the Keplerian specific angular momentum 
$\sqrt{GMR{_A}}$ close to Alfv\'en or magnetospheric radius. For stellar wind accretion, 
the specific angular momentum of the captured material is considered to be significantly low, 
thus leading the material to fall straight on the neutron star magnetosphere without any disk-formation 
\citep{Ho1989, Frank2002}. 

Two types of quasi-spherical wind accretion onto the neutron stars are expected 
\citep{Shakura2012, Shakura2018}. The supersonic quasi-spherical accretion, also  
known as the Bondi-Hoyle-Littleton accretion, occurs when the accreting shocked 
plasma rapidly cools down and falls supersonically towards the magnetosphere. The 
subsonic quasi-spherical accretion is possible in slow rotating magnetized neutron 
star with X-ray luminosity $<$4$\times$10$^{36}$~\lumcgs where the accreted plasma 
remains hot at the magnetospheric boundary, \citep{Shakura2012}. In the latter case, 
the material forms a hot quasi-static shell around the rotating neutron star magnetosphere 
in process of setting subsonically. This shell mediates the angular momentum transfer 
to- or from the neutron star by large-scale convective turbulent structures. The spin-up 
or spin-down of the neutron stars are expected to observe depending on the sign of 
angular velocity difference between the magnetosphere and the accreting material. 
The model can successfully explain the secular spin variation of the pulsar 
along with any irregular short-term frequency fluctuations.

\begin{figure}
\centering
\includegraphics[height=2.6in, width=3.3in, angle=0]{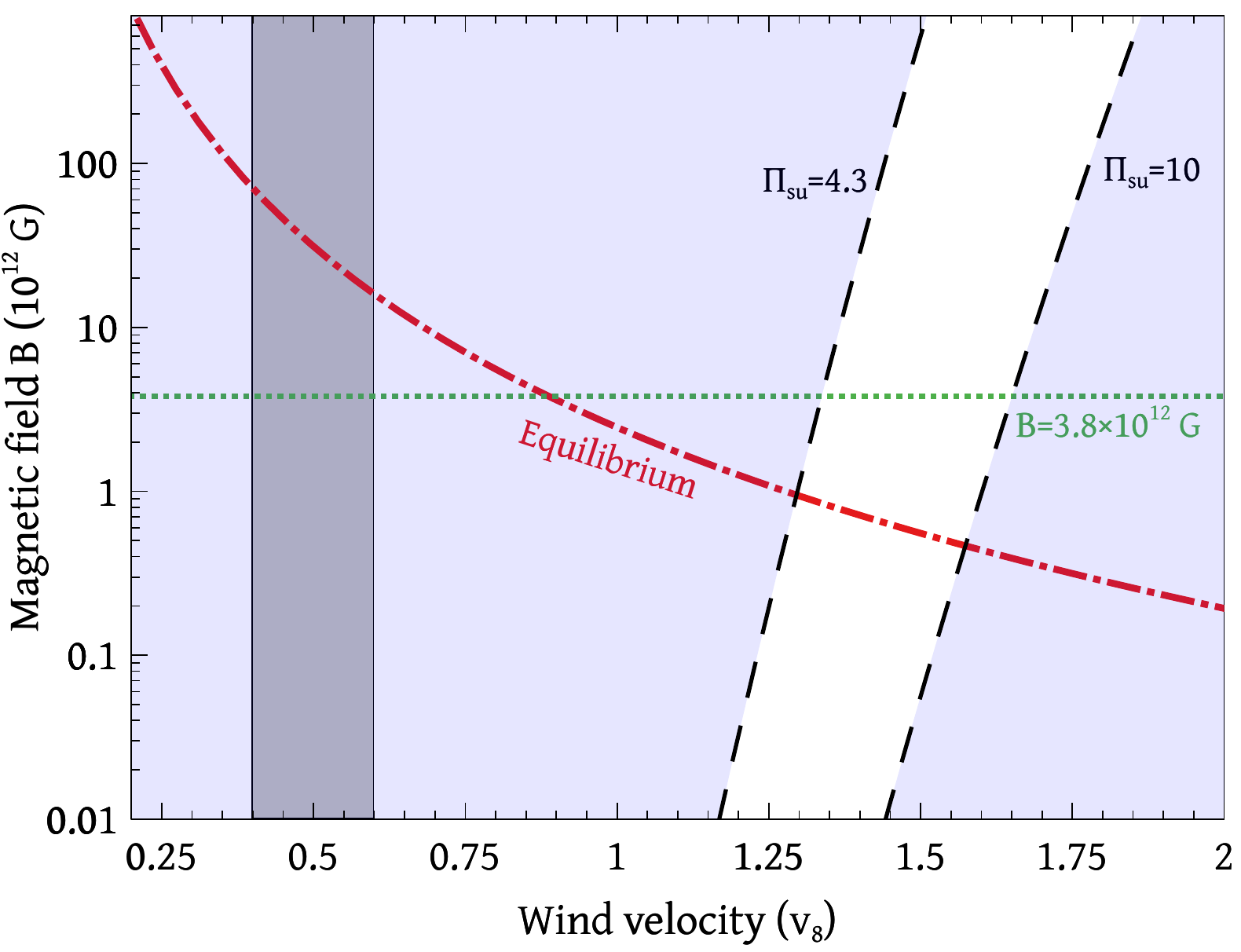}
\caption{Magnetic field vs. wind velocity relation for \source. The dashed lines (black) indicates 
the variation between magnetic field and wind velocity (in a unit of 10$^8$ cm~s$^{-1}$) for a given parameter $\Pi_{su}$
of the settling accretion theory at 4.3 and 10 (white region). The  dash-dot red curve 
shows the magnetic field in case of spin equilibrium. Observed wind velocity of the optical 
companion is also shown in the grey range. The dashed horizontal line (green) 
corresponds to the magnetic field estimated by a tentative estimation of cyclotron line 
\citep{Jaisawal2013}}. 
\label{fig-mf}
\end{figure}

We apply the quasi-spherical settling accretion theory estimating the magnetic field 
of \source from the applied torque. According to the theory (\citealt{Shakura2012}; 
Equation~4 of \citealt{Postnov2015MNRAS4461013P}), the spin-up rate is 
\begin{equation}\label{2}
 \dot\omega_{su}\simeq 10^{-9}[Hz~d^{-1}]~\Pi_{su}~\mu_{30}^{1/11}
v_{8}^{-4}\left(\frac{P_{b}}{10~d}\right)^{-1}\dot M_{16}^{7/11} 
\end{equation}   
here the spin rate $\dot\omega_{su}$ is measured to be 2.56$\times$10$^{-9}$[Hz~d$^{-1}$] 
for \source. $\Pi_{su}$ is a system independent dimensionless parameter of the theory ranging 
between 4.3 and 10 \citep{Shakura2012, Postnov2015MNRAS4461013P, Shakura2018}. The dipole magnetic 
moment of the neutron star is $\mu_{30}$=$\mu$/10$^{30}$[G~cm$^3$] and is related to the magnetic 
field (B) by expression $\mu$=BR$^3$/2 (the neutron star radius R is assumed to be 10~km). The 
parameter v$_8$ is the stellar wind velocity in the unit of 10$^8$[cm~s$^{-1}$]. The mass 
accretion rate $\dot M_{16}$=$\dot M$/10$^{16}$[g~s$^{-1}$] is estimated to be $\dot M_{16}$=0.69 
for the source luminosity of 1.24$\times$10$^{36}$~\lumcgs at 20\% accretion efficiency. 
The orbital period P$_b$ of the system is 4.4~d. In this regime, the magnetic field strongly 
depends on the wind velocity that has been estimated to be 500$\pm$100~\kms for \source 
\citep{Martnez2015}. We found that the settling model requires approximately three times 
higher wind velocity for \source between 1300--1500~\kms to get a magnetic field in a range 
of 10$^{12}$--10$^{14}$~G (white region in Figure~\ref{fig-mf}) for a given value of $\Pi_{su}$. 
We can not adequately determine the magnetic field using Equation~\ref{2} of the settling 
accretion theory.

We notice that the observed source luminosity is stable at $\sim$10$^{36}$~\lumcgs 
during each X-ray observations so far. Despite the variation in spin period, 
considering an extreme case, we may assume the pulsar is accreting close to 
its equilibrium spin period that is  : 
\begin{equation}\label{3}
 P_{eq}\simeq 940[s]~\mu_{30}^{12/11}
v_{8}^{4}\left(\frac{P_{b}}{10~d}\right)\dot M_{16}^{-4/11} 
\end{equation}   

The above equation is adopted from \citet{Shakura2012} \& \citet{Postnov2015MNRAS4461013P}. 
Considering the wind velocity between 400--600~\kms and the observed mass accretion rate, 
the spin equilibrium magnetic field can be estimated to be $>$10$^{13}$~G (red curve in 
Figure~\ref{fig-mf}). The inferred value is within a factor of five times higher than the 
polar magnetic field ($\approx$7.6$\times$10$^{12}$~G -- twice the equatorial field) inferred from 
the tentative detection of cyclotron line  by \citet{Jaisawal2013} which has not yet been re-confirmed. The measurement appears to be reasonable, given 
many theoretical uncertainties, as well as the fact that the pulsar is not strictly at 
equilibrium due to changes in the spin frequency.

\subsection{Pulse profile and spectroscopy}

We examine the emission geometry of the pulsar using the \nustar and \astrosat 
observations in 2015 July and 2017 July, respectively. This is done by studying the pulse 
profiles that strongly depend on the energy. We observed a complex, broad profile 
in soft X-rays that evolves into a narrow emission peak followed by plateau like 
structure in hard X-rays. Almost similar energy evolution and pulse morphology were 
detected in the earlier studies \citep{Furst2011,Furst2012,Jaisawal2013}. The 
existence of anomalous pulse profiles at several epochs advocates the persistent 
emission geometry of the pulsar. Even so, the marginal difference in the profile 
morphology exists that possibly led by the local variation in column density and 
distribution of the stellar wind across the binary orbit. 

The emission from one or both the poles of the neutron star produces
a relatively complex pulse profile depending on the geometrical and
gravitation effects \citep{Kraus1995, Bulik2003, Lutovinov2009, Sasaki2012}. The absorption of 
soft X-ray photons emitted from the accretion column by the accreting 
material complicates the profile further. Sometimes this material influences 
the direct pulsar emission so much that multiple absorption dips are observed at 
definite pulse phases of the pulsar. Such dips have been frequently seen in the 
pulse profiles of Be/X-ray binary pulsars during X-ray outbursts (see e.g., 
\citealt{Maitra2012,Naik2013,Jaisawal2016,Ferrigno2016,Epili2017,Gupta2018}). In the present 
study, the accretion columns at both the poles of the pulsar contribute as the 
pulse profiles below 10 keV are seen to be double-peaked. We observed a narrow 
emission peak in a 0.1--0.2 phase range of the pulse profile at higher energies. 
The detection of high pulse fraction (about 50\%) above 20~keV is expected from a 
narrow emission region (Figure~\ref{pp}).

The energy spectrum of pulsars originates from thermal and bulk Comptonization physical 
processes in the accretion column \citep{Becker2007}. The emission from \source can be 
described by an absorbed HECut model as seen during the 2015 and 2017 observations. The 
spectral parameters are consistent during these observations, except the value of the 
column density. \nh is detected to be high 18$\times$10$^{22}$~cm$^{-2}$ during the 
\astrosat observation, covering 0.87--0.05 orbital phase range. We also observed a 
relatively lower value of column density 10$\times$10$^{22}$~cm$^{-2}$ using \nustar 
and XRT data obtained in 0.215--0.437 orbital phase range. These variations are in 
agreement with the findings by \citet{Levine2004}. The authors also found a strong 
increase in \nh below 0.12 orbital phase. The orbital changes in the column density 
can be explained by considering the movement of the neutron star through the inhomogeneously
distributed stellar wind. To probe the wind distribution further, we performed phase-resolved
spectroscopy using \nustar data in 2015. We detected a phase variation in the \nh which 
likely affects the X-ray emission from the accretion column. The \nh is found to be three 
times higher in the plateau region of the pulse profile as opposed to the value observed 
at 0.1--0.2 pulse phase. A narrow emission peak is clearly present 0.1--0.2 pulse phase 
range when the column density is low.

The cyclotron absorption line is understood to originate from the resonant scattering 
of photons with the electrons in the quantized Landau levels \citep{Meszaros1992}. These 
features are observed in the 10--100 keV energy range of pulsar spectrum corresponding to 
a magnetic field of 10$^{12}$--10$^{13}$ Gauss, based on the 12-B-12 rule 
\citep{Jaisawal2017,Staubert2019}. A cyclotron line was tentatively 
observed at 44 keV in \source using the \suzaku data \citep{Jaisawal2013}. 
However, in the present study, any signature of a cyclotron absorption line in 
phase-averaged and phase-resolved spectroscopy of the pulsar spectrum is not 
detected. The spin-equilibrium magnetic field can match the polar B-strength
provided by cyclotron line within a factor of 5, possibly due to the 
theoretical uncertainties in the calculation. Future longer exposure 
observations with instruments like \nustar, \astrosat, and the proposed 
{\em STROBE-X} mission \citep{Ray2019} could confirm the detection/ 
non-detection of a cyclotron absorption feature in \source.

\

\subsection*{ACKNOWLEDGMENTS}
We thank the anonymous referee for constructive suggestions on the paper.
This research has made use of data obtained through 
HEASARC Online Service, provided by the NASA/GSFC, in support of NASA 
High Energy Astrophysics Programs. This work used the NuSTAR Data Analysis 
Software ({\tt NuSTARDAS}) jointly developed by the ASI
Science Data Center (ASDC, Italy) and the California Institute of Technology (USA).
This publication uses the data from the \astrosat mission of the 
Indian Space Research Organisation (ISRO), archived at the Indian 
Space Science Data Centre (ISSDC). We thank members of SXT, LAXPC, and CZTI 
instrument teams for their contribution to
the development of the instruments and analysis software. 
We also acknowledge the contributions of the \astrosat project team at ISAC and IUCAA. 
S.N. and N.K. acknowledge the supports from Physical Research Laboratory 
which is funded by the Department of Space, Government of India. 
This research has made use of ISIS functions (ISISscripts) provided 
by ECAP/Remeis observatory and MIT (http://www.sternwarte.uni-erlangen.de/isis/)

\section*{DATA AVAILABILITY}
We used archival data of {\it NuSTAR}, {\it Swift},  and \astrosat observatories for this work.

\bibliographystyle{mnras}
\bibliography{references.bib}

\begin{thebibliography}{}
\makeatletter
\relax
\def\mn@urlcharsother{\let\do\@makeother \do\$\do\&\do\#\do\^\do\_\do\%\do\~}
\def\mn@doi{\begingroup\mn@urlcharsother \@ifnextchar [ {\mn@doi@}
  {\mn@doi@[]}}
\def\mn@doi@[#1]#2{\def\@tempa{#1}\ifx\@tempa\@empty \href
  {http://dx.doi.org/#2} {doi:#2}\else \href {http://dx.doi.org/#2} {#1}\fi
  \endgroup}
\def\mn@eprint#1#2{\mn@eprint@#1:#2::\@nil}
\def\mn@eprint@arXiv#1{\href {http://arxiv.org/abs/#1} {{\tt arXiv:#1}}}
\def\mn@eprint@dblp#1{\href {http://dblp.uni-trier.de/rec/bibtex/#1.xml}
  {dblp:#1}}
\def\mn@eprint@#1:#2:#3:#4\@nil{\def\@tempa {#1}\def\@tempb {#2}\def\@tempc
  {#3}\ifx \@tempc \@empty \let \@tempc \@tempb \let \@tempb \@tempa \fi \ifx
  \@tempb \@empty \def\@tempb {arXiv}\fi \@ifundefined
  {mn@eprint@\@tempb}{\@tempb:\@tempc}{\expandafter \expandafter \csname
  mn@eprint@\@tempb\endcsname \expandafter{\@tempc}}}

\bibitem[\protect\citeauthoryear{{Agrawal}}{{Agrawal}}{2006}]{Agrawal2006}
{Agrawal} P.~C.,  2006, \mn@doi [Advances in Space Research]
  {10.1016/j.asr.2006.03.038}, \href
  {https://ui.adsabs.harvard.edu/abs/2006AdSpR..38.2989A} {38, 2989}

\bibitem[\protect\citeauthoryear{{Agrawal} et~al.,}{{Agrawal}
  et~al.}{2017}]{Agrawal2017}
{Agrawal} P.~C.,  et~al., 2017, \mn@doi [Journal of Astrophysics and Astronomy]
  {10.1007/s12036-017-9451-z}, \href
  {https://ui.adsabs.harvard.edu/abs/2017JApA...38...30A} {38, 30}

\bibitem[\protect\citeauthoryear{{Antia} et~al.,}{{Antia}
  et~al.}{2017}]{Antia2017}
{Antia} H.~M.,  et~al., 2017, \mn@doi [\apjs] {10.3847/1538-4365/aa7a0e}, \href
  {https://ui.adsabs.harvard.edu/abs/2017ApJS..231...10A} {231, 10}

\bibitem[\protect\citeauthoryear{{Arnaud}}{{Arnaud}}{1996}]{Arnaud1996}
{Arnaud} K.~A.,  1996, in {Jacoby} G.~H.,  {Barnes} J.,  eds,  Astronomical
  Society of the Pacific Conference Series Vol. 101, Astronomical Data Analysis
  Software and Systems V. p.~17

\bibitem[\protect\citeauthoryear{{Basko} \& {Sunyaev}}{{Basko} \&
  {Sunyaev}}{1975}]{Basko1975}
{Basko} M.~M.,  {Sunyaev} R.~A.,  1975, Astronomy and Astrophysics, \href
  {https://ui.adsabs.harvard.edu/abs/1975A&A....42..311B} {42, 311}

\bibitem[\protect\citeauthoryear{{Becker} \& {Wolff}}{{Becker} \&
  {Wolff}}{2007}]{Becker2007}
{Becker} P.~A.,  {Wolff} M.~T.,  2007, \mn@doi [\apj] {10.1086/509108}, \href
  {https://ui.adsabs.harvard.edu/abs/2007ApJ...654..435B} {654, 435}

\bibitem[\protect\citeauthoryear{{Bulik}, {Gondek-Rosi{\'n}ska}, {Santangelo},
  {Mihara}, {Finger}  \& {Cemeljic}}{{Bulik} et~al.}{2003}]{Bulik2003}
{Bulik} T.,  {Gondek-Rosi{\'n}ska} D.,  {Santangelo} A.,  {Mihara} T.,
  {Finger} M.,   {Cemeljic} M.,  2003, \mn@doi [\aap]
  {10.1051/0004-6361:20030555}, \href
  {https://ui.adsabs.harvard.edu/abs/2003A&A...404.1023B} {404, 1023}

\bibitem[\protect\citeauthoryear{{Caballero} \& {Wilms}}{{Caballero} \&
  {Wilms}}{2012}]{Caballero2012}
{Caballero} I.,  {Wilms} J.,  2012, Memorie della Societa Astronomica Italiana,
  \href {https://ui.adsabs.harvard.edu/abs/2012MmSAI..83..230C} {83, 230}

\bibitem[\protect\citeauthoryear{{Corbet} \& {Krimm}}{{Corbet} \&
  {Krimm}}{2013}]{Corbet2013}
{Corbet} R. H.~D.,  {Krimm} H.~A.,  2013, \mn@doi [\apj]
  {10.1088/0004-637X/778/1/45}, \href
  {https://ui.adsabs.harvard.edu/abs/2013ApJ...778...45C} {778, 45}

\bibitem[\protect\citeauthoryear{{Deeter}, {Boynton}, {Lamb}  \&
  {Zylstra}}{{Deeter} et~al.}{1989}]{Deeter1989}
{Deeter} J.~E.,  {Boynton} P.~E.,  {Lamb} F.~K.,   {Zylstra} G.,  1989, \mn@doi
  [\apj] {10.1086/167017}, \href
  {https://ui.adsabs.harvard.edu/abs/1989ApJ...336..376D} {336, 376}

\bibitem[\protect\citeauthoryear{{Epili}, {Naik}, {Jaisawal}  \&
  {Gupta}}{{Epili} et~al.}{2017}]{Epili2017}
{Epili} P.,  {Naik} S.,  {Jaisawal} G.~K.,   {Gupta} S.,  2017, \mn@doi
  [\mnras] {10.1093/mnras/stx2247}, \href
  {http://adsabs.harvard.edu/abs/2017MNRAS.472.3455E} {472, 3455}

\bibitem[\protect\citeauthoryear{{Evans} et~al.,}{{Evans}
  et~al.}{2009}]{Evans2009}
{Evans} P.~A.,  et~al., 2009, \mn@doi [\mnras]
  {10.1111/j.1365-2966.2009.14913.x}, \href
  {http://adsabs.harvard.edu/abs/2009MNRAS.397.1177E} {397, 1177}

\bibitem[\protect\citeauthoryear{{Farinelli}, {Ceccobello}, {Romano}  \&
  {Titarchuk}}{{Farinelli} et~al.}{2012}]{Farinelli2012}
{Farinelli} R.,  {Ceccobello} C.,  {Romano} P.,   {Titarchuk} L.,  2012,
  \mn@doi [\aap] {10.1051/0004-6361/201118008}, \href
  {https://ui.adsabs.harvard.edu/abs/2012A&A...538A..67F} {538, A67}

\bibitem[\protect\citeauthoryear{{Ferrigno}, {Pjanka}, {Bozzo}, {Klochkov},
  {Ducci}  \& {Zdziarski}}{{Ferrigno} et~al.}{2016}]{Ferrigno2016}
{Ferrigno} C.,  {Pjanka} P.,  {Bozzo} E.,  {Klochkov} D.,  {Ducci} L.,
  {Zdziarski} A.~A.,  2016, \mn@doi [\aap] {10.1051/0004-6361/201527837}, \href
  {https://ui.adsabs.harvard.edu/abs/2016A&A...593A.105F} {593, A105}

\bibitem[\protect\citeauthoryear{{Forman}, {Jones}, {Cominsky}, {Julien},
  {Murray}, {Peters}, {Tananbaum}  \& {Giacconi}}{{Forman}
  et~al.}{1978}]{Forman1978}
{Forman} W.,  {Jones} C.,  {Cominsky} L.,  {Julien} P.,  {Murray} S.,  {Peters}
  G.,  {Tananbaum} H.,   {Giacconi} R.,  1978, \mn@doi [The Astrophysical
  Journal Supplement Series] {10.1086/190561}, \href
  {https://ui.adsabs.harvard.edu/abs/1978ApJS...38..357F} {38, 357}

\bibitem[\protect\citeauthoryear{{Frank}, {King}  \& {Raine}}{{Frank}
  et~al.}{2002}]{Frank2002}
{Frank} J.,  {King} A.,   {Raine} D.~J.,  2002, {Accretion Power in
  Astrophysics: Third Edition}

\bibitem[\protect\citeauthoryear{{F{\"u}rst}, {Kreykenbohm}, {Suchy},
  {Barrag{\'a}n}, {Wilms}, {Rothschild}  \& {Pottschmidt}}{{F{\"u}rst}
  et~al.}{2011}]{Furst2011}
{F{\"u}rst} F.,  {Kreykenbohm} I.,  {Suchy} S.,  {Barrag{\'a}n} L.,  {Wilms}
  J.,  {Rothschild} R.~E.,   {Pottschmidt} K.,  2011, \mn@doi [\aap]
  {10.1051/0004-6361/201015636}, \href
  {https://ui.adsabs.harvard.edu/abs/2011A&A...525A..73F} {525, A73}

\bibitem[\protect\citeauthoryear{{F{\"u}rst}, {Pottschmidt}, {Kreykenbohm},
  {M{\"u}ller}, {K{\"u}hnel}, {Wilms}  \& {Rothschild}}{{F{\"u}rst}
  et~al.}{2012}]{Furst2012}
{F{\"u}rst} F.,  {Pottschmidt} K.,  {Kreykenbohm} I.,  {M{\"u}ller} S.,
  {K{\"u}hnel} M.,  {Wilms} J.,   {Rothschild} R.~E.,  2012, \mn@doi [\aap]
  {10.1051/0004-6361/201219845}, \href
  {https://ui.adsabs.harvard.edu/abs/2012A&A...547A...2F} {547, A2}

\bibitem[\protect\citeauthoryear{{Ghosh} \& {Lamb}}{{Ghosh} \&
  {Lamb}}{1979}]{Ghosh1979}
{Ghosh} P.,  {Lamb} F.~K.,  1979, \mn@doi [\apj] {10.1086/157498}, \href
  {https://ui.adsabs.harvard.edu/abs/1979ApJ...234..296G} {234, 296}

\bibitem[\protect\citeauthoryear{{Giacconi}, {Murray}, {Gursky}, {Kellogg},
  {Schreier}, {Matilsky}, {Koch}  \& {Tananbaum}}{{Giacconi}
  et~al.}{1974}]{Giacconi1974}
{Giacconi} R.,  {Murray} S.,  {Gursky} H.,  {Kellogg} E.,  {Schreier} E.,
  {Matilsky} T.,  {Koch} D.,   {Tananbaum} H.,  1974, \mn@doi [The
  Astrophysical Journal Supplement Series] {10.1086/190288}, \href
  {https://ui.adsabs.harvard.edu/abs/1974ApJS...27...37G} {27, 37}

\bibitem[\protect\citeauthoryear{{Gupta}, {Naik}, {Jaisawal}  \&
  {Epili}}{{Gupta} et~al.}{2018}]{Gupta2018}
{Gupta} S.,  {Naik} S.,  {Jaisawal} G.~K.,   {Epili} P.~R.,  2018, \mn@doi
  [\mnras] {10.1093/mnras/sty1804}, \href
  {https://ui.adsabs.harvard.edu/abs/2018MNRAS.479.5612G} {479, 5612}

\bibitem[\protect\citeauthoryear{{HI4PI Collaboration} et~al.,}{{HI4PI
  Collaboration} et~al.}{2016}]{HI4PI2016}
{HI4PI Collaboration} et~al., 2016, \mn@doi [\aap]
  {10.1051/0004-6361/201629178}, \href
  {https://ui.adsabs.harvard.edu/abs/2016A&A...594A.116H} {594, A116}

\bibitem[\protect\citeauthoryear{{Harrison} et~al.,}{{Harrison}
  et~al.}{2013}]{Harrison2013}
{Harrison} F.~A.,  et~al., 2013, \mn@doi [\apj] {10.1088/0004-637X/770/2/103},
  \href {http://adsabs.harvard.edu/abs/2013ApJ...770..103H} {770, 103}

\bibitem[\protect\citeauthoryear{{Ho}, {Taam}, {Fryxell}, {Matsuda}  \&
  {Koide}}{{Ho} et~al.}{1989}]{Ho1989}
{Ho} C.,  {Taam} R.~E.,  {Fryxell} B.~A.,  {Matsuda} T.,   {Koide} H.,  1989,
  \mn@doi [\mnras] {10.1093/mnras/238.4.1447}, \href
  {https://ui.adsabs.harvard.edu/abs/1989MNRAS.238.1447H} {238, 1447}

\bibitem[\protect\citeauthoryear{{Jaisawal} \& {Naik}}{{Jaisawal} \&
  {Naik}}{2017}]{Jaisawal2017}
{Jaisawal} G.~K.,  {Naik} S.,  2017, in {Serino} M.,  {Shidatsu} M.,  {Iwakiri}
  W.,   {Mihara} T.,  eds, 7 years of MAXI: monitoring X-ray Transients, held
  5-7 December 2016 at RIKEN. p.~153 (\mn@eprint {arXiv} {1705.05536})

\bibitem[\protect\citeauthoryear{{Jaisawal}, {Naik}  \& {Paul}}{{Jaisawal}
  et~al.}{2013}]{Jaisawal2013}
{Jaisawal} G.~K.,  {Naik} S.,   {Paul} B.,  2013, \mn@doi [\apj]
  {10.1088/0004-637X/779/1/54}, \href
  {https://ui.adsabs.harvard.edu/abs/2013ApJ...779...54J} {779, 54}

\bibitem[\protect\citeauthoryear{{Jaisawal}, {Naik}  \& {Epili}}{{Jaisawal}
  et~al.}{2016}]{Jaisawal2016}
{Jaisawal} G.~K.,  {Naik} S.,   {Epili} P.,  2016, \mn@doi [\mnras]
  {10.1093/mnras/stw085}, \href
  {http://adsabs.harvard.edu/abs/2016MNRAS.457.2749J} {457, 2749}

\bibitem[\protect\citeauthoryear{{Jaisawal}, {Naik}  \& {Chenevez}}{{Jaisawal}
  et~al.}{2018}]{Jaisawal+18}
{Jaisawal} G.~K.,  {Naik} S.,   {Chenevez} J.,  2018, \mn@doi [\mnras]
  {10.1093/mnras/stx3082}, \href
  {http://adsabs.harvard.edu/abs/2018MNRAS.474.4432J} {474, 4432}

\bibitem[\protect\citeauthoryear{{Jaisawal} et~al.,}{{Jaisawal}
  et~al.}{2019}]{Jaisawal2019}
{Jaisawal} G.~K.,  et~al., 2019, \mn@doi [\apj] {10.3847/1538-4357/ab4595},
  \href {https://ui.adsabs.harvard.edu/abs/2019ApJ...885...18J} {885, 18}

\bibitem[\protect\citeauthoryear{{Kraus}, {Nollert}, {Ruder}  \&
  {Riffert}}{{Kraus} et~al.}{1995}]{Kraus1995}
{Kraus} U.,  {Nollert} H.~P.,  {Ruder} H.,   {Riffert} H.,  1995, \mn@doi
  [\apj] {10.1086/176182}, \href
  {https://ui.adsabs.harvard.edu/abs/1995ApJ...450..763K} {450, 763}

\bibitem[\protect\citeauthoryear{{Leahy}}{{Leahy}}{1987}]{Leahy1987}
{Leahy} D.~A.,  1987, \aap, \href
  {https://ui.adsabs.harvard.edu/abs/1987A&A...180..275L} {180, 275}

\bibitem[\protect\citeauthoryear{{Levine}, {Rappaport}, {Remillard}  \&
  {Savcheva}}{{Levine} et~al.}{2004}]{Levine2004}
{Levine} A.~M.,  {Rappaport} S.,  {Remillard} R.,   {Savcheva} A.,  2004,
  \mn@doi [The Astrophysical Journal] {10.1086/425567}, \href
  {https://ui.adsabs.harvard.edu/abs/2004ApJ...617.1284L} {617, 1284}

\bibitem[\protect\citeauthoryear{{Lewin}, {van Paradijs}  \& {van den
  Heuvel}}{{Lewin} et~al.}{1997}]{Lewin1997}
{Lewin} W.~H.~G.,  {van Paradijs} J.,   {van den Heuvel} E.~P.~J.,  1997,
  {X-ray Binaries}

\bibitem[\protect\citeauthoryear{{Lutovinov} \& {Tsygankov}}{{Lutovinov} \&
  {Tsygankov}}{2009}]{Lutovinov2009}
{Lutovinov} A.~A.,  {Tsygankov} S.~S.,  2009, \mn@doi [Astronomy Letters]
  {10.1134/S1063773709070019}, \href
  {https://ui.adsabs.harvard.edu/abs/2009AstL...35..433L} {35, 433}

\bibitem[\protect\citeauthoryear{{Maitra}, {Paul}  \& {Naik}}{{Maitra}
  et~al.}{2012}]{Maitra2012}
{Maitra} C.,  {Paul} B.,   {Naik} S.,  2012, \mn@doi [\mnras]
  {10.1111/j.1365-2966.2011.20196.x}, \href
  {https://ui.adsabs.harvard.edu/abs/2012MNRAS.420.2307M} {420, 2307}

\bibitem[\protect\citeauthoryear{{Mart{\'\i}nez-N{\'u}{\~n}ez}, {Sander},
  {G{\'\i}menez-Garc{\'\i}a}, {G{\'o}nzalez-Gal{\'a}n}, {Torrej{\'o}n},
  {G{\'o}nzalez-Fern{\'a}ndez}  \& {Hamann}}{{Mart{\'\i}nez-N{\'u}{\~n}ez}
  et~al.}{2015}]{Martnez2015}
{Mart{\'\i}nez-N{\'u}{\~n}ez} S.,  {Sander} A.,  {G{\'\i}menez-Garc{\'\i}a} A.,
   {G{\'o}nzalez-Gal{\'a}n} A.,  {Torrej{\'o}n} J.~M.,
  {G{\'o}nzalez-Fern{\'a}ndez} C.,   {Hamann} W.~R.,  2015, \mn@doi [\aap]
  {10.1051/0004-6361/201424823}, \href
  {https://ui.adsabs.harvard.edu/abs/2015A&A...578A.107M} {578, A107}

\bibitem[\protect\citeauthoryear{{Mart{\'\i}nez-N{\'u}{\~n}ez}
  et~al.,}{{Mart{\'\i}nez-N{\'u}{\~n}ez} et~al.}{2017}]{Mart2017}
{Mart{\'\i}nez-N{\'u}{\~n}ez} S.,  et~al., 2017, \mn@doi [\ssr]
  {10.1007/s11214-017-0340-1}, \href
  {https://ui.adsabs.harvard.edu/abs/2017SSRv..212...59M} {212, 59}

\bibitem[\protect\citeauthoryear{{Meszaros}}{{Meszaros}}{1992}]{Meszaros1992}
{Meszaros} P.,  1992, {High-energy radiation from magnetized neutron stars}

\bibitem[\protect\citeauthoryear{{Misra} et~al.,}{{Misra}
  et~al.}{2017}]{Misra2017}
{Misra} R.,  et~al., 2017, \mn@doi [\apj] {10.3847/1538-4357/835/2/195}, \href
  {https://ui.adsabs.harvard.edu/abs/2017ApJ...835..195M} {835, 195}

\bibitem[\protect\citeauthoryear{{Morel} \& {Grosdidier}}{{Morel} \&
  {Grosdidier}}{2005}]{Morel2005}
{Morel} T.,  {Grosdidier} Y.,  2005, \mn@doi [\mnras]
  {10.1111/j.1365-2966.2004.08488.x}, \href
  {https://ui.adsabs.harvard.edu/abs/2005MNRAS.356..665M} {356, 665}

\bibitem[\protect\citeauthoryear{{Nagase}}{{Nagase}}{1989}]{Nagase1989}
{Nagase} F.,  1989, \pasj, \href
  {https://ui.adsabs.harvard.edu/abs/1989PASJ...41....1N} {41, 1}

\bibitem[\protect\citeauthoryear{{Naik}, {Paul}  \& {Ali}}{{Naik}
  et~al.}{2011}]{Naik2011}
{Naik} S.,  {Paul} B.,   {Ali} Z.,  2011, \mn@doi [\apj]
  {10.1088/0004-637X/737/2/79}, \href
  {https://ui.adsabs.harvard.edu/abs/2011ApJ...737...79N} {737, 79}

\bibitem[\protect\citeauthoryear{{Naik}, {Maitra}, {Jaisawal}  \&
  {Paul}}{{Naik} et~al.}{2013}]{Naik2013}
{Naik} S.,  {Maitra} C.,  {Jaisawal} G.~K.,   {Paul} B.,  2013, \mn@doi [\apj]
  {10.1088/0004-637X/764/2/158}, \href
  {https://ui.adsabs.harvard.edu/abs/2013ApJ...764..158N} {764, 158}

\bibitem[\protect\citeauthoryear{{Odaka}, {Khangulyan}, {Tanaka}, {Watanabe},
  {Takahashi}  \& {Makishima}}{{Odaka} et~al.}{2013}]{Odaka2013}
{Odaka} H.,  {Khangulyan} D.,  {Tanaka} Y.~T.,  {Watanabe} S.,  {Takahashi} T.,
    {Makishima} K.,  2013, \mn@doi [\apj] {10.1088/0004-637X/767/1/70}, \href
  {https://ui.adsabs.harvard.edu/abs/2013ApJ...767...70O} {767, 70}

\bibitem[\protect\citeauthoryear{{Paul} \& {Naik}}{{Paul} \&
  {Naik}}{2011}]{Paul2011}
{Paul} B.,  {Naik} S.,  2011, Bulletin of the Astronomical Society of India,
  \href {https://ui.adsabs.harvard.edu/abs/2011BASI...39..429P} {39, 429}

\bibitem[\protect\citeauthoryear{{Porter} \& {Rivinius}}{{Porter} \&
  {Rivinius}}{2003}]{Porter2003}
{Porter} J.~M.,  {Rivinius} T.,  2003, \mn@doi [\pasp] {10.1086/378307}, \href
  {https://ui.adsabs.harvard.edu/abs/2003PASP..115.1153P} {115, 1153}

\bibitem[\protect\citeauthoryear{{Postnov}, {Mironov}, {Lutovinov}, {Shakura},
  {Kochetkova}  \& {Tsygankov}}{{Postnov}
  et~al.}{2015}]{Postnov2015MNRAS4461013P}
{Postnov} K.~A.,  {Mironov} A.~I.,  {Lutovinov} A.~A.,  {Shakura} N.~I.,
  {Kochetkova} A.~Y.,   {Tsygankov} S.~S.,  2015, \mn@doi [\mnras]
  {10.1093/mnras/stu2155}, \href
  {https://ui.adsabs.harvard.edu/abs/2015MNRAS.446.1013P} {446, 1013}

\bibitem[\protect\citeauthoryear{{Ramadevi} et~al.,}{{Ramadevi}
  et~al.}{2018}]{Ramadevi2018}
{Ramadevi} M.~C.,  et~al., 2018, \mn@doi [Journal of Astrophysics and
  Astronomy] {10.1007/s12036-017-9506-1}, \href
  {https://ui.adsabs.harvard.edu/abs/2018JApA...39...11R} {39, 11}

\bibitem[\protect\citeauthoryear{{Rao}, {Bhattacharya}, {Bhalerao}, {Vadawale}
  \& {Sreekumar}}{{Rao} et~al.}{2017}]{Rao2017}
{Rao} A.~R.,  {Bhattacharya} D.,  {Bhalerao} V.~B.,  {Vadawale} S.~V.,
  {Sreekumar} S.,  2017, Current Science, \href
  {https://ui.adsabs.harvard.edu/abs/2017CSci..113..595R} {113, 595}

\bibitem[\protect\citeauthoryear{{Ray} et~al.,}{{Ray} et~al.}{2011}]{Ray2011}
{Ray} P.~S.,  et~al., 2011, \mn@doi [\apjs] {10.1088/0067-0049/194/2/17}, \href
  {https://ui.adsabs.harvard.edu/abs/2011ApJS..194...17R} {194, 17}

\bibitem[\protect\citeauthoryear{{Ray} et~al.,}{{Ray} et~al.}{2019a}]{Ray2019}
{Ray} P.~S.,  et~al., 2019a, arXiv e-prints, \href
  {https://ui.adsabs.harvard.edu/abs/2019arXiv190303035R} {p. arXiv:1903.03035}

\bibitem[\protect\citeauthoryear{{Ray} et~al.,}{{Ray}
  et~al.}{2019b}]{2019ApJ...879..130R}
{Ray} P.~S.,  et~al., 2019b, \mn@doi [\apj] {10.3847/1538-4357/ab24d8}, \href
  {https://ui.adsabs.harvard.edu/abs/2019ApJ...879..130R} {879, 130}

\bibitem[\protect\citeauthoryear{{Reig}}{{Reig}}{2011}]{Reig2011}
{Reig} P.,  2011, \mn@doi [\apss] {10.1007/s10509-010-0575-8}, \href
  {https://ui.adsabs.harvard.edu/abs/2011Ap&SS.332....1R} {332, 1}

\bibitem[\protect\citeauthoryear{{Reig} \& {Nespoli}}{{Reig} \&
  {Nespoli}}{2013}]{Reig2013}
{Reig} P.,  {Nespoli} E.,  2013, \mn@doi [\aap] {10.1051/0004-6361/201219806},
  \href {https://ui.adsabs.harvard.edu/abs/2013A&A...551A...1R} {551, A1}

\bibitem[\protect\citeauthoryear{{Revnivtsev} \& {Mereghetti}}{{Revnivtsev} \&
  {Mereghetti}}{2015}]{Revnivtsev2015}
{Revnivtsev} M.,  {Mereghetti} S.,  2015, \mn@doi [\ssr]
  {10.1007/s11214-014-0123-x}, \href
  {https://ui.adsabs.harvard.edu/abs/2015SSRv..191..293R} {191, 293}

\bibitem[\protect\citeauthoryear{{Sasaki}, {M{\"u}ller}, {Kraus}, {Ferrigno}
  \& {Santangelo}}{{Sasaki} et~al.}{2012}]{Sasaki2012}
{Sasaki} M.,  {M{\"u}ller} D.,  {Kraus} U.,  {Ferrigno} C.,   {Santangelo} A.,
  2012, \mn@doi [\aap] {10.1051/0004-6361/201016304}, \href
  {https://ui.adsabs.harvard.edu/abs/2012A&A...540A..35S} {540, A35}

\bibitem[\protect\citeauthoryear{{Shakura}, {Postnov}, {Kochetkova}  \&
  {Hjalmarsdotter}}{{Shakura} et~al.}{2012}]{Shakura2012}
{Shakura} N.,  {Postnov} K.,  {Kochetkova} A.,   {Hjalmarsdotter} L.,  2012,
  \mn@doi [\mnras] {10.1111/j.1365-2966.2011.20026.x}, \href
  {https://ui.adsabs.harvard.edu/abs/2012MNRAS.420..216S} {420, 216}

\bibitem[\protect\citeauthoryear{{Shakura}, {Postnov}, {Kochetkova}  \&
  {Hjalmarsdotter}}{{Shakura} et~al.}{2018}]{Shakura2018}
{Shakura} N.,  {Postnov} K.,  {Kochetkova} A.~r.,   {Hjalmarsdotter} L.,  2018,
  {Quasi-Spherical Subsonic Accretion onto Magnetized Neutron Stars}.
p.~331, \mn@doi{10.1007/978-3-319-93009-1_7}

\bibitem[\protect\citeauthoryear{{Singh} et~al.,}{{Singh}
  et~al.}{2017}]{Singh2017}
{Singh} K.~P.,  et~al., 2017, \mn@doi [Journal of Astrophysics and Astronomy]
  {10.1007/s12036-017-9448-7}, \href
  {https://ui.adsabs.harvard.edu/abs/2017JApA...38...29S} {38, 29}

\bibitem[\protect\citeauthoryear{{Staubert} et~al.,}{{Staubert}
  et~al.}{2019}]{Staubert2019}
{Staubert} R.,  et~al., 2019, \mn@doi [\aap] {10.1051/0004-6361/201834479},
  \href {https://ui.adsabs.harvard.edu/abs/2019A&A...622A..61S} {622, A61}

\bibitem[\protect\citeauthoryear{{Tandon} et~al.,}{{Tandon}
  et~al.}{2017}]{Tandon2017}
{Tandon} S.~N.,  et~al., 2017, \mn@doi [\aj] {10.3847/1538-3881/aa8451}, \href
  {https://ui.adsabs.harvard.edu/abs/2017AJ....154..128T} {154, 128}

\bibitem[\protect\citeauthoryear{{Tauris} \& {van den Heuvel}}{{Tauris} \& {van
  den Heuvel}}{2006}]{Tauris2006}
{Tauris} T.~M.,  {van den Heuvel} E.~P.~J.,  2006, {Formation and evolution of
  compact stellar X-ray sources}.
pp 623--665

\bibitem[\protect\citeauthoryear{{Torrej{\'o}n}, {Schulz}, {Nowak}  \&
  {Kallman}}{{Torrej{\'o}n} et~al.}{2010}]{Torrejon2010}
{Torrej{\'o}n} J.~M.,  {Schulz} N.~S.,  {Nowak} M.~A.,   {Kallman} T.~R.,
  2010, \mn@doi [\apj] {10.1088/0004-637X/715/2/947}, \href
  {https://ui.adsabs.harvard.edu/abs/2010ApJ...715..947T} {715, 947}

\bibitem[\protect\citeauthoryear{{Tsygankov}, {Lutovinov}, {Doroshenko},
  {Mushtukov}, {Suleimanov}  \& {Poutanen}}{{Tsygankov}
  et~al.}{2016}]{Tsygankov2016}
{Tsygankov} S.~S.,  {Lutovinov} A.~A.,  {Doroshenko} V.,  {Mushtukov} A.~A.,
  {Suleimanov} V.,   {Poutanen} J.,  2016, \mn@doi [\aap]
  {10.1051/0004-6361/201628236}, \href
  {https://ui.adsabs.harvard.edu/abs/2016A&A...593A..16T} {593, A16}

\bibitem[\protect\citeauthoryear{{Vasilopoulos}, {Petropoulou}, {Koliopanos},
  {Ray}, {Bailyn}, {Haberl}  \& {Gendreau}}{{Vasilopoulos}
  et~al.}{2019}]{2019MNRAS.488.5225V}
{Vasilopoulos} G.,  {Petropoulou} M.,  {Koliopanos} F.,  {Ray} P.~S.,  {Bailyn}
  C.~B.,  {Haberl} F.,   {Gendreau} K.,  2019, \mn@doi [\mnras]
  {10.1093/mnras/stz2045}, \href
  {https://ui.adsabs.harvard.edu/abs/2019MNRAS.488.5225V} {488, 5225}

\bibitem[\protect\citeauthoryear{{Vasilopoulos} et~al.,}{{Vasilopoulos}
  et~al.}{2020}]{2020MNRAS.494.5350V}
{Vasilopoulos} G.,  et~al., 2020, \mn@doi [\mnras] {10.1093/mnras/staa991},
  \href {https://ui.adsabs.harvard.edu/abs/2020MNRAS.494.5350V} {494, 5350}

\bibitem[\protect\citeauthoryear{{Walter}, {Lutovinov}, {Bozzo}  \&
  {Tsygankov}}{{Walter} et~al.}{2015}]{Walter2015}
{Walter} R.,  {Lutovinov} A.~A.,  {Bozzo} E.,   {Tsygankov} S.~S.,  2015,
  \mn@doi [Astronomy and Astrophysics Review] {10.1007/s00159-015-0082-6},
  \href {https://ui.adsabs.harvard.edu/abs/2015A&ARv..23....2W} {23, 2}

\bibitem[\protect\citeauthoryear{{Wen}, {Remillard}  \& {Bradt}}{{Wen}
  et~al.}{2000}]{Wen2000}
{Wen} L.,  {Remillard} R.~A.,   {Bradt} H.~V.,  2000, \mn@doi [The
  Astrophysical Journal] {10.1086/308604}, \href
  {https://ui.adsabs.harvard.edu/abs/2000ApJ...532.1119W} {532, 1119}

\bibitem[\protect\citeauthoryear{{White}, {Swank}  \& {Holt}}{{White}
  et~al.}{1983}]{White1983}
{White} N.~E.,  {Swank} J.~H.,   {Holt} S.~S.,  1983, \mn@doi [\apj]
  {10.1086/161162}, \href {http://adsabs.harvard.edu/abs/1983ApJ...270..711W}
  {270, 711}

\bibitem[\protect\citeauthoryear{{Wilms}, {Allen}  \& {McCray}}{{Wilms}
  et~al.}{2000}]{Wilms2000}
{Wilms} J.,  {Allen} A.,   {McCray} R.,  2000, \mn@doi [\apj] {10.1086/317016},
  \href {http://adsabs.harvard.edu/abs/2000ApJ...542..914W} {542, 914}

\bibitem[\protect\citeauthoryear{{Wilson-Hodge} et~al.,}{{Wilson-Hodge}
  et~al.}{2018}]{Wilson2018}
{Wilson-Hodge} C.~A.,  et~al., 2018, \mn@doi [\apj] {10.3847/1538-4357/aace60},
  \href {http://adsabs.harvard.edu/abs/2018ApJ...863....9W} {863, 9}

\makeatother
\end{thebibliography}

\bsp	
\label{lastpage}

\end{document}